\documentclass[%
 aps, prx,
 amsmath,amssymb,
 reprint,%
superscriptaddress
]{revtex4-2}

\usepackage{graphicx}
\usepackage{bm}
\usepackage{fixme}
\usepackage{hyperref}
\usepackage{kbordermatrix}
\usepackage[utf8]{inputenc}
\usepackage[T1]{fontenc}
\usepackage{mathptmx}
\usepackage{lipsum}
\usepackage{amsmath}
\usepackage{physics}
\usepackage{xparse}
\usepackage{bbm}
\usepackage{xcolor}
\usepackage{url}

\graphicspath{{Pictures/}}

\newcommand{\mytitile}{Light-dressing of a 
diatomic superconducting artificial molecule}

\begin{document}
	\preprint{AIP/123-QED}
	
	\title[\mytitile]{\mytitile\\~}
	\author{G. P. Fedorov}
	\email{gleb.fedorov@phystech.edu}
	
	\affiliation{ 
		Russian Quantum Center, Skolkovo village, Russia
	}%
	\affiliation{ 
		Moscow Institute of Physics and Technology, Dolgoprundiy, Russia
	}
	\affiliation{
		National University of Science and Technology MISIS, Moscow, Russia
	}%
	\author{V. B. Yursa}
	\affiliation{Skolkovo Institute of Science 
		and Technology, Moscow, Russian Federation}%

\affiliation{ 
	Moscow Institute of Physics and Technology, 
	Dolgoprundiy, Russia
}%
	
	\author{A. E. Efimov}
	
	\affiliation{ 
		Moscow Institute of Physics and Technology, Dolgoprundiy, Russia
	}%
	
	\author{K. I. Shiianov}

	\affiliation{ 
		Moscow Institute of Physics and Technology, Dolgoprundiy, Russia
	}%

	\author{A. Yu. Dmitriev}
	\affiliation{ 
	Moscow Institute of Physics and Technology, Dolgoprundiy, Russia
	}%

	\author{I. A. Rodionov}
	\affiliation{FMN Laboratory, Bauman Moscow 
	State Technical University, Moscow, Russia}
	\affiliation{Dukhov Automatics Research 
	Institute, (VNIIA), Moscow, Russia}

	\author{A. A. Dobronosova}
	\affiliation{FMN Laboratory, Bauman Moscow 
	State Technical University, Moscow, Russia}
	\affiliation{Dukhov Automatics Research 
	Institute, (VNIIA), Moscow, Russia}

	\author{D. O. Moskalev}
	\affiliation{FMN Laboratory, Bauman Moscow 
	State Technical University, Moscow, Russia}

	\author{A. A. Pishchimova}
	\affiliation{FMN Laboratory, Bauman Moscow 
	State Technical University, Moscow, Russia}
	\affiliation{Dukhov Automatics Research 
	Institute, (VNIIA), Moscow, Russia}
	
	\author{E. I. Malevannaya}
	\affiliation{FMN Laboratory, Bauman Moscow 
		State Technical University, Moscow, Russia}
	\affiliation{Dukhov Automatics Research 
		Institute, (VNIIA), Moscow, Russia}

	\author{O. V. Astafiev}
	\affiliation{Skolkovo Institute of Science 
		and Technology, Moscow, Russian Federation}
	\affiliation{ 
		Moscow Institute of Physics and Technology, 
		Dolgoprundiy, Russia
	}
	\affiliation{Physics Department, Royal 
	Holloway, University of London, Egham, Surrey 
	TW20 0EX, United Kingdom}
\affiliation{National Physical Laboratory, Teddington, TW11 0LW, United Kingdom}

	\date{\today}

	\begin{abstract}
	In this work, we irradiate a superconducting 
	artificial molecule composed of two coupled tunable transmons with 
	microwave light while monitoring its state 
	via joint dispersive readout. Performing high-power spectroscopy, we observe and identify a variety of single- and multiphoton transitions. We also find that at certain 
	fluxes, the measured spectrum of the system deviates significantly from the 
	solution of the stationary Schrödinger equation with no
	driving. We reproduce these unusual 
	spectral features by solving numerically the full master equation for a steady-state and attribute them to an Autler-Townes-like effect
	in which a single tone is simultaneously dressing 
	the system and probing the transitions 
	between new eigenstates. We show that it is possible to find analytically the exact frequencies at which the satellite spectral lines appear by solving self-consistent equations in the rotating frame. Our approach 
	agrees well with both the experiment and the numerical 
	simulation.
	\end{abstract}
	
	\maketitle
\section{Introduction}

Over the past twenty years, superconducting 
artificial atoms (SAAs) were used in numerous 
experiments in a compelling demonstration of the 
validity of fundamental quantum mechanical 
laws \cite{you2011atomic,Qopt_1300refs}. Their Hamiltonians can be pre-designed 
and engineered which makes them a particularly 
versatile tool for studies in quantum optics, and high controllability of their parameters 
allows direct observation of novel physical effects previously inaccessible for natural 
systems. 

One of the most prominent milestones that 
superconducting quantum circuits have reached so 
far is the strong coupling with light in {circuit 
QED} \cite{wallraff2004strong, 
chiorescu2004coherent} when the relaxation and decoherence rates 
appear smaller than the Rabi frequency. 
Currently, they are surpassing all other 
implementations of strong coupling in terms of 
coherence \cite{forn2019ultrastrong}. 
However, in sharp contrast with natural atoms and 
molecules, SAAs do not even require confined 
radiation to implement strong coupling with 
light: they may be coupled unprecedently strongly 
to free-propagating electromagnetic waves in 
on-chip waveguides \cite{astafiev2010resonance} 
without using cavities at all. In this case, the Rabi frequency may reach 50\% of the driven transition 
frequency \cite{deng2015observation} which is even 
beyond strong coupling 
regime \cite{forn2019ultrastrong}. To correctly 
describe the atomic behavior in these 
conditions, the so-called \textit{dressed atom 
approach} \cite{cohen1998atom} is employed: 
the radiation has to be directly included in the 
Hamiltonian of the system and affects its level structure.

Thus far, there have been many experiments with intense driving fields revealing  dressing effects in on-chip quantum optics with artificial 
atoms \cite{baur2009measurement, 
sillanpaa2009autler, astafiev2010resonance, 
novikov2013autler, suri2013observation, 
koshino2013observation, 
braumuller2015multiphoton, peng2018vacuum, 
gasparinetti2019two}. In all these works, light 
dressing of the atom manifests itself through 
Mollow triplets or Autler-Townes (A-T) splittings of 
different kinds. However, despite the recent 
successes in control of large arrays of 
interacting SAAs \cite{Song574, ye2019propagation, 
arute2019quantum}, there were no experiments concerning the behavior
of similar composite structures under a strong drive. While there were studies on dressing 
of multi-atomic systems in a 
cavity \cite{fink2009dressed, 
macha2014implementation, shulga2017observation, 
yang2018probing}, dressing by an external free field is no less attractive since its 
frequency may be easily tuned into resonance with any transition of the system. Moreover, the drive amplitude is also easier to tune than the coupling strength for cavity-dressed systems.

In this work, we study a pair of strongly coupled 
artificial atoms: a superconducting artificial 
molecule \cite{kou2017fluxonium} (SAM). We use two 
Xmon-type 
transmons \cite{koch2007charge, barends2013coherent} interacting with 
each other both through a cavity 
bus \cite{majer2007coupling} and a direct capacitance. Microwave radiation 
is applied to this system through an on-chip 
coplanar waveguide antenna while its state can be 
monitored by joint dispersive readout using the 
same cavity \cite{chow2010detecting}. Examining our high-resolution spectroscopic data, we
find that strong interaction with microwaves not 
only results in a rich variety of multiphoton transitions of 
various orders between SAM states, but also 
significantly modifies its level structure. Even 
in a simple diatomic molecule, this leads to 
complex Autler-Townes-like effects involving 
single- and multiphoton  transitions that can 
only be explained in the dressed picture. Even though the A-T splittings have been 
investigated before in a wide range of quantum 
systems (including natural 
molecules \cite{tamarat1995pump, 
ahmed2012autler}), we find qualitatively new 
spectral manifestations of light dressing when 
SAAs are irradiated unequally. We could not find any reports of similar effects in the 
previous studies of SAAs under a strong driving field. Prior works 
either involved just a single atom \cite{baur2009measurement, 
	sillanpaa2009autler, astafiev2010resonance, 
	novikov2013autler, 
	koshino2013observation, 
	braumuller2015multiphoton, peng2018vacuum, 
	gasparinetti2019two} or 
demonstrated only the standard spectral 
signatures known from the quantum optics \cite{suri2013observation} (see 
Appendix \ref{sec:3-level-at}). Moreover, previous spectroscopic experiments with coupled transmons were either done at low powers and resolved only the most prominent single-photon transitions \cite{majer2007coupling, filipp2011multimode}, or used simultaneous excitation at two distinct frequencies to reach higher energy levels \cite{dicarlo2009demonstration}, or did not study the spectral data with necessary resolution \cite{kounalakis2018tuneable}, or used non-tunable transmons \cite{poletto2012entanglement}. In contrast, we now put a tunable system in a new regime of intense driving which allows us to discover and quantify both experimentally and theoretically its novel unexplored behaviors. Likewise, we could not find reports of such effects in natural molecules, 
which could in principle be observed there; most probably, this is caused by reduced controllability and 
coherence compared to superconducting quantum devices. 

We believe that our results are valuable to the domain of molecular physics and quantum optics beyond just superconducting Josephson systems since the reported effects are possible to find in any kind of light-matter interaction. In similar conditions, they will emerge for any diatomic molecule regardless of its nature, and modification of the molecular spectra using light is now a topic of active research \cite{hertzog2019strong}. In this regard, we note that our theoretical framework can be employed in the analysis of similar effects in the future. Besides, we consider this experiment important for superconducting quantum 
computing: one should take the observed behaviors into account and control carefully the 
drive power (for instance, as we will show, the 
bSWAP gate \cite{poletto2012entanglement} may be 
directly affected by light dressing). Finally, the high-power excitation may be applied to directly obtain information about higher energy levels of the system using minimum equipment; this approach can facilitate the scaling of control electronics for superconducting quantum processors (see, i.e. \cite{hornibrook2015cryogenic}).

The manuscript consists of four main parts and an 
Appendix. Section I is this introduction; Section 
II is devoted to the approaches used in 
our study; Section III contains the results of 
our experimental and theoretical research, 
including numerical simulations and analytical 
analysis; finally, in Section IV we make a 
conclusion of our work and discuss future 
prospects. The Appendix contains the 
details of the theory that we use and additional information about the sample and the measurement techniques.

\section{Methods}\label{sec:methods}

\subsection{Device design and control}

\begin{figure}
	\includegraphics[width=\linewidth]{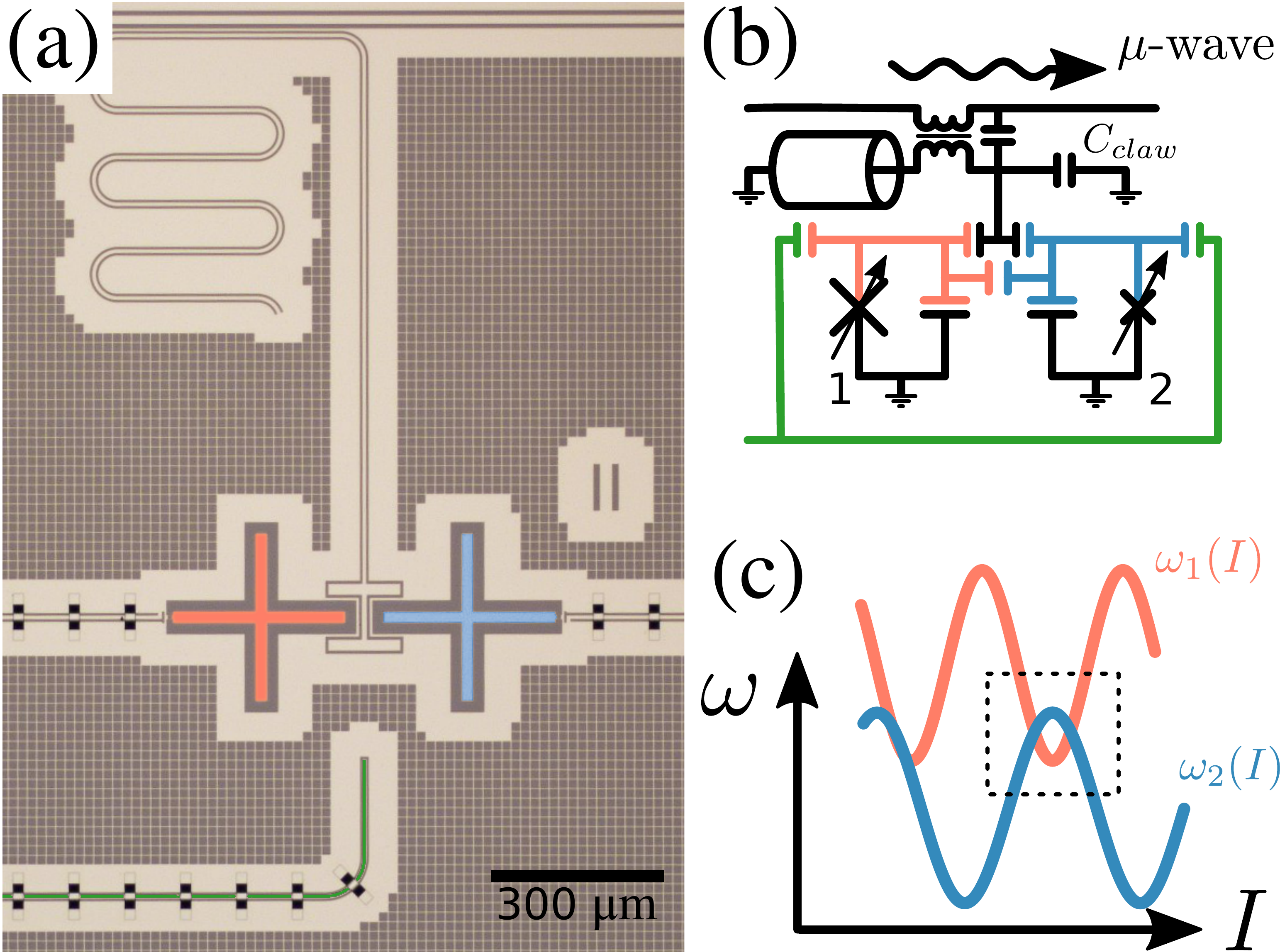}
	\caption{\textbf{(a)} Optical image of 
		the device (false colored). Two transmons 
		(orange, 1 and blue, 2) are coupled 
		capacitively to a $\lambda/4$ coplanar 
		resonator. Frequency control lines come from 
		both sides, and from below a waveguide for 
		microwave excitation  is connected (green). 
		\textbf{(b)} The equivalent electrical 
		circuit. Tunable Josephson junctions are	SQUIDs with magnetic flux control. 
		\textbf{(c)} A sketch of the transmon frequencies $\omega_1(I)$ 
		and $\omega_2(I)$ depending on the current 
		$I$ in an external coil when correctly 
		aligned by the individual flux control lines. 
		In this work, we focus on the area inside the 
		dashed rectangle.}
	\label{fig:experiment}
\end{figure}

We have designed the SAM as a pair of tunable Xmon SAAs with asymmetric 
SQUIDs \cite{hutchings2017tunable}. They are coupled 
to a single notch-type 
$\lambda/4$ resonator \cite{probst2015efficient} ($f_r = 7.34$ GHz, $Q_e \approx 1900,\ Q_l \approx 1100$) which 
serves for the dispersive readout of their states \cite{chow2010detecting}. In \autoref{fig:experiment}~(a), 
the optical image of the device is shown, presenting the physical layout of the components. The 
resonator is connected to a 
coplanar waveguide through which the readout is 
performed. At its open end, it is coupled to 
the transmons by a dual 
``claw'' coupler \cite{barends2013coherent}. Flux 
lines allowing independent control of the 
transmon frequencies are coming from the sides, and the excitation waveguide from 
below directs the microwave signal towards the SAM (green). 
In \autoref{fig:experiment}~(b), the equivalent 
electrical circuit of the device is shown. The 
resonator is inductively and capacitively coupled to the transmission line. Here we simplify the distributed coupling down to lumped elements, as in \cite{khalil2012analysis}, even though there is a more rigorous approach for this case \cite{besedin2018quality}. The transmons 1 and 2 are false-colored orange and blue, respectively, and their SQUIDs are 
represented as tunable Josephson junctions. One can note that our design gives rise to two types of coupling in the SAM. The first is the dispersive virtual photon exchange through the multimode cavity\cite{majer2007coupling, filipp2011multimode} and the second is the direct coupling via a mutual capacitance. We find that both of them contribute noticeably to the observed effective coupling strength, but with opposite signs (details can be found in Appendix \ref{sec:full_model}). In 
\autoref{fig:experiment}~(c) we show 
schematically the transmon frequencies 
versus the electric current $I$ which we apply 
to an external coil wound around the sample 
holder. Since the effective junction of the 
transmon 1 is larger, its main transition (orange) lies 
higher in frequency than the one of the 
second transmon (blue). Using individual 
flux-control lines, it is possible to align the 
SAAs so that the lower sweet-spot of the  
transmon 1 is just below the upper sweet-spot of 
the transmon 2 (see the dashed rectangle,  
\autoref{fig:experiment}~(c)). As we will show in the following, this configuration 
is convenient 
to track the energies of highly excited levels 
via multi-photon transitions in a single 
spectroscopic scan. Additionally, the transmons 
are better protected from the flux noise near their sweet spots.

\subsection{Quantum-mechanical description}

A single transmon SAA can be regarded as an 
oscillator with a quartic perturbation describing 
the leading-order anharmonicity \cite{koch2007charge, yan2018tunable}. Therefore, in 
the main text we do not use the charge and the phase 
operators and write down its Hamiltonian using 
only the annihilation operator $\hat b$:
\begin{equation}
\hat{{H}}_{tr}/\hbar = \omega \hat 
b^{\dagger}\hat b +\frac{1}{2}\alpha \hat 
b^{\dagger}\hat b(\hat b^{\dagger}\hat b-1),
\label{eq:h1tr}
\end{equation}
where $\omega$ is the 
$\ket{0}\rightarrow\ket{1}$, or fundamental, 
transition frequency and $\alpha$ is the 
anharmonicity. By applying a
magnetic flux to the SQUID (either via 
an individual on-chip line or via an external 
coil) it is possible to 
directly  control $\omega$ \cite{koch2007charge}. In our modeling, we 
take into account the three lowest states of the 
transmon ($\ket{0},\ \ket{1}$ and $\ket{2}$).

\autoref{eq:h1tr} describes the SAA without 
driving. To model a monochromatic 
microwave signal of frequency $\omega_d$ applied 
through a capacitively coupled transmission line, 
the following driving term should be included in 
the Hamiltonian:
\begin{equation}
\hat H_{d} = \hbar \Omega (\hat b+\hat b^{\dagger}) \cos\omega_d t,
\end{equation}
where $\Omega$ is the driving amplitude coinciding with the frequency of the Rabi oscillations between $\ket{0}$ and $\ket{1}$.

Next, we assemble the model for two coupled 
transmons with the corresponding annihilation 
operators $\hat b$ and $\hat c$, the fundamental 
frequencies $\omega_{1,2}$ and anharmonicities 
$\alpha_{1,2}$. The corresponding Hamiltonian of 
the SAM contains two terms representing each 
transmon, two terms representing the interaction 
of the transmons with the driving fields at  $\omega_d^{(1,2)}$, and the 
transmon-transmon interaction term:
\begin{equation}\label{Hsystem}
\hat H = \hat H_{tr}^{(1)}+\hat H_{tr}^{(2)}+\hat H_{d}^{(1)}+\hat H_{d}^{(2)}+\hat H_{int},
\end{equation}
where the superscripts numerate the transmons and $\hat H_{int} = \hbar J (\hat b +\hat 
b^\dag)(\hat c+\hat c^{\dagger})$. 
Strictly speaking, $J = J(\omega_1, \omega_2)$ 
depends on the transmon 
frequencies \cite{koch2007charge}, but we take $J$ 
to be a constant due to its negligible variation for our range of frequencies (see also Appendix \ref{sec:full_model} for details).

For brevity, the SAM Hamiltonian without driving 
terms and the corresponding eigenenergies will be 
referred below as ``unperturbed''. Since we use three levels for each transmon, there is a total of nine 
basis states of the SAM $\ket{i}\otimes \ket{j} = 
\ket{ij}$, where $i$ and $j$ show the number of 
excitations in the first and the second transmon, respectively.

In the following, we will also transform 
\autoref{Hsystem} into the frame rotating with 
both drives by an operator
\begin{equation}
\hat R = \exp[-i t (\omega_d^{(1)}
b^{\dagger}b+\omega_d^{(2)} 
c^{\dagger}c)],\label{eq:R}
\end{equation}
arriving at
\begin{equation}
\hat H_R = \hat R^{\dagger}\hat H \hat R -	 
{i}\hat R^{\dagger}\partial_t \hat 
R.\label{eq:rotation}
\end{equation}
After the transformation and application of the RWA
\begin{equation}
\begin{aligned}
	\omega_{1,2} &\rightarrow \Delta_{1,2} = \omega_{1,2} - \omega_d^{(1,2)},\\
	\hat H_{int} &\rightarrow \hbar J \left[\hat 
	b^\dag \hat c e^{it(\omega_d^{(1)} - \omega_d^{(2)})} 
	+ \hat b \hat c^\dag e^{-it(\omega_d^{(1)} - 
	\omega_d^{(2)})}\right],\\
	\hat H_{d}^{(1)} &\rightarrow \frac{\hbar \Omega_1}{2}(\hat b  + \hat b^\dag),\ 	\hat H_{d}^{(2)} \rightarrow \frac{\hbar \Omega_2}{2}(\hat c  + \hat c^\dag).
\end{aligned}
\label{eq:RWA}
\end{equation}
Note that if the transmons are driven at the same frequency, the RWA Hamiltonian is time independent. Below we will use the symbol  $
\omega_d = \omega_d^{(1)} = 
\omega_d^{(2)}$ to denote that common frequency of both drives.

Besides the unitary evolution, we also model the incoherent processes of relaxation 
and dephasing for each transmon using the Lindblad equation with the 
following collapse 
operators \cite{bishop2010circuit}:
\begin{equation}\
\begin{split}
\hat{{O}}_{\gamma}^{(1)} = \sqrt{\gamma^{(1)}}\, \hat b,\ 
\hat{{O}}_{\phi}^{(1)} = \sqrt{\gamma_{\phi}^{(1)}}\, 
\hat b^\dag \hat b,\\
\hat{{O}}_{\gamma}^{(2)} = \sqrt{\gamma^{(2)}}\, \hat c,\ 
\hat{{O}}_{\phi}^{(2)} = \sqrt{\gamma_{ \phi}^{(2)}}\, 
\hat c^\dag \hat c,
\end{split}
\end{equation}
where $\gamma^{(1,2)}$ are the individual 
relaxation rates, and $\gamma_{\phi}^{(1,2)}$ are 
the pure dephasing rates. As one can see, the 
collapse operators are in a separable form, i.e. 
acting only upon a single transmon each, which is 
a valid approach as long as the coupling strength $J \ll \omega_{1,2}$ 
 \cite{beaudoin2011dissipation}. 
Therefore, the complete evolution equation for 
the system density matrix $\hat \rho$ is		
\begin{equation}
\partial_t \hat \rho_{(R)} = \frac{i}{\hbar}[\hat 
\rho, \hat H_{(R)}] + \sum_{\substack{\alpha = {\gamma, \phi},\\ i=1,2}} 
\mathcal{D}[\hat{O}_{\alpha}^{(i)}] \hat \rho_{(R)} 
= \mathcal{L}\hat\rho_{(R)}, \label{eq:master}
\end{equation}
where $\mathcal{D}[\hat{{O}}]\hat \rho = 
\hat{{O}} \hat \rho \hat{{O}}^\dag - 
\frac{1}{2}\{ \hat{{O}}^\dag \hat{{O}}, \hat 
\rho\}$ and $\mathcal{L}$ is the Liouville 
superoperator, or the Liouvillian; $_{(R)}$ 
denotes if the Hamiltonian and the corresponding 
solution density matrix are in the rotating frame 
with RWA. In this work, we do not alter the 
dissipator terms when changing the reference 
frame despite that it may not be correct in 
general \cite{shavit2019bridging}.

We summarize the parameters of the SAM model in \autoref{tab:parameters}. The coherence times $T_1 = 1/\gamma$, $T_2^{*} = 1/(\gamma/2 + \gamma_{\phi})$ were measured independently in the lower and the upper sweet spots for transmons 1 and 2, respectively. The remaining parameters were extracted from the fits of the unperturbed model to the observed spectral lines; this procedure will be discussed in more detail in Section \ref{sec:level1}. The underlying electrical parameters of the transmons are the Josephson energies $E^{(1)}_{J, 
\sum}/h = 24.3$ GHz, $E^{(2)}_{J,\sum}/h = 18.3$ GHz,  the charging energies $E^{(1,2)}_C/h = 220$ MHz, and the SQUID asymmetries $d^{(1,2)} = 0.7$ (all these parameters are defined as in \cite{koch2007charge})

\begin{table}
	\begin{ruledtabular}
		\begin{tabular}{rll}
			Parameter & Transmon 1  & Transmon 2\\\hline
			$\omega/2\pi$ & 5.12 - 6.30  GHz & 4.00 - 5.45 GHz\\
			$\alpha/2\pi$ & -220 MHz & -220 MHz \\
			$T_1$  & 6.82 $\mu$s &  4.41 $\mu$s \\
			$T_2^*$  & 5.14 $\mu$s  &  3.33 $\mu$s\\\hline
			$J/2\pi$  &\multicolumn{2}{c}{8.69 MHz} 
		\end{tabular}
	\end{ruledtabular}
	\caption{SAM model parameters. Transmons are only 
		different in the frequency tuning range and 
		coherence times measured in the lower sweet 
		spot for the 1\textsuperscript{st} one and in 
		the higher sweet spot for the 
		2\textsuperscript{nd}. The coupling strength 
		$J$ depends on the transmon frequencies and 
		is specified here for $\omega_1/2\pi = 
		\omega_2/2\pi = 5.32$ GHz.}
	\label{tab:parameters}
\end{table}	

The readout resonator is not included explicitly in the 
above model since it does not 
affect the dynamics of the SAM in the dispersive regime. To model the 
readout, we use an ad-hoc measurement 
operator $\hat M(f_p)$ that can be obtained by 
finding the transmission $S^{(ij)}_{21}(f_p)$ ($f_p$ is 
the probe frequency near the resonator frequency) through 
the sample after preparing various states $\ket{ij}$ of the 
SAM. However, in this work we use a simpler method which is to 
calculate $S^{(ij)}_{21}(f_p)$ via offsetting the 
experimental resonance curve $S^{(00)}_{21}(f_p)$ measured while the 
SAM is in the ground state by the corresponding dispersive 
shifts $\chi_{ij}$ \cite{filipp2009two, chow2010detecting}. We calculate  $\chi_{ij}$ with the parameters that can be found in Appendix \ref{sec:full_model} according to \cite{koch2007charge}, Eq. D3. Finally, the observable value for any state $\hat 
\rho$ is calculated as $S_{21}^{sim}(f_p) = 
\Tr[\hat M(f_p) \hat \rho]$. 

\subsection{Numerical solution in \textit{qutip}}

Numerical simulations are necessary for studying 
\autoref{eq:master} since it does not have an 
analytical solution. We have been using the 
\textit{qutip} \cite{johansson2013qutip} package to simulate the 
dynamics and to find the steady state of the 
system for various parameter combinations of the 
Liouvillian. The source code for the simulations can be found on GitHub \footnote{\url{https://github.com/vdrhtc/Examples/tree/master/transmon-simulations/two_transmons}}. 

Two distinct modes of simulation were used. The first one is for the Liouvillians 
that do not explicitly depend on time. In this 
case, the steady state $\hat \rho_{ss}$ of the 
system should be calculated from the set of 
linear equations obtained from \autoref{eq:master}
\begin{equation}
\partial_t \hat \rho = 0  \Rightarrow \mathcal{L} 
\hat \rho = 0
\label{eq:steady}
\end{equation}
This equation is solved with the qutip's 
\textit{steadystate} 
function \footnote{\url{http://qutip.org/docs/4.0.2/guide/guide-steady.html}}.
This method is applicable when the driving 
frequency for both transmons is the same.

The second mode is required when it is not 
possible to avoid the time-dependence of the 
Liouvillian or if one wants to solve the master 
equation in the laboratory frame. For example, 
when the transmons are excited at 
different frequencies ($\omega_d^{(1)} - \omega_d^{(2)} = 
\delta \neq 0$), from \autoref{eq:RWA} we find 
that even in the doubly-rotating frame, $\hat 
H_{int}$ is oscillating, and it is not possible 
to simply drop this term in RWA because it is 
inherent for the SAM. In this case, to find the 
steady state of the system one can employ the 
functions \textit{propagator} and 
\textit{propagator\_steadystate} of \textit{qutip}. The 
propagator is a completely positive map 
$\Lambda(t_1, t_0): \hat \rho(t_0) \rightarrow 
\hat \rho(t_1)$ describing the time evolution of 
the system density matrix; for 
\autoref{eq:master}, it is defined as
\begin{equation}
\Lambda(t_1, t_0) = \mathcal{T} \exp [\int_{t_0}^{t_1} \mathcal L(\tau) d\tau],
\label{eq:propagator}
\end{equation}
where $\mathcal T$ is the time-ordering 
superoperator. Since the Liouvillian is periodic 
with a period  $T = 2\pi/\delta$, it is 
possible to calculate the steady state as the 
eigenvector $\hat \rho_{ss}$ of the single-period 
propagator $\Lambda(T, 0)$ corresponding to its 
largest eigenvalue  \cite{dittrich1998quantum, 
rivas2012open}. Upon infinitely many applications 
of $\Lambda$
\[
\lim_{n\to \infty} \left[\Lambda(T, 0)\right]^n \hat \rho = \Lambda(nT, 0) \hat \rho \to \hat \rho_{ss}.
\]

\section{Results}

\subsection{\label{sec:level1} Spectroscopy: experiment and numerical simulation}

\begin{figure*}
	
	\centering
	\includegraphics[width=\linewidth]{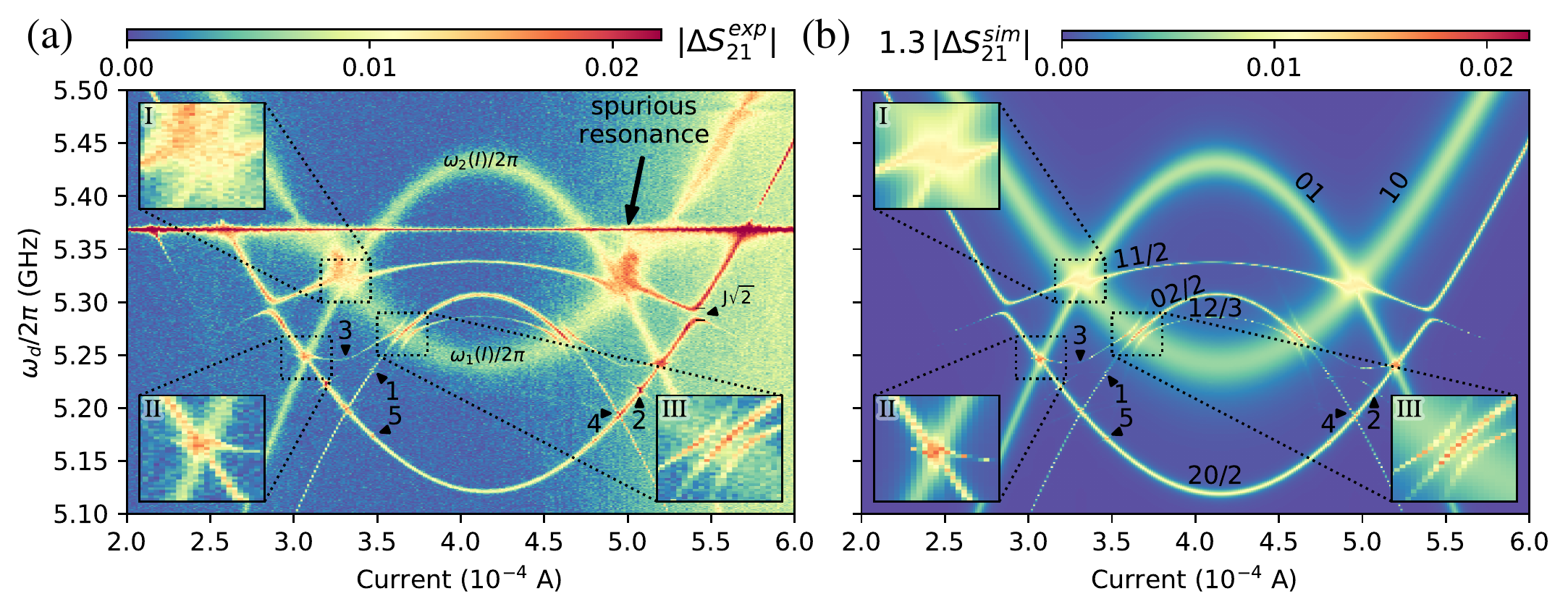}
	
	\includegraphics[width=.495\linewidth]{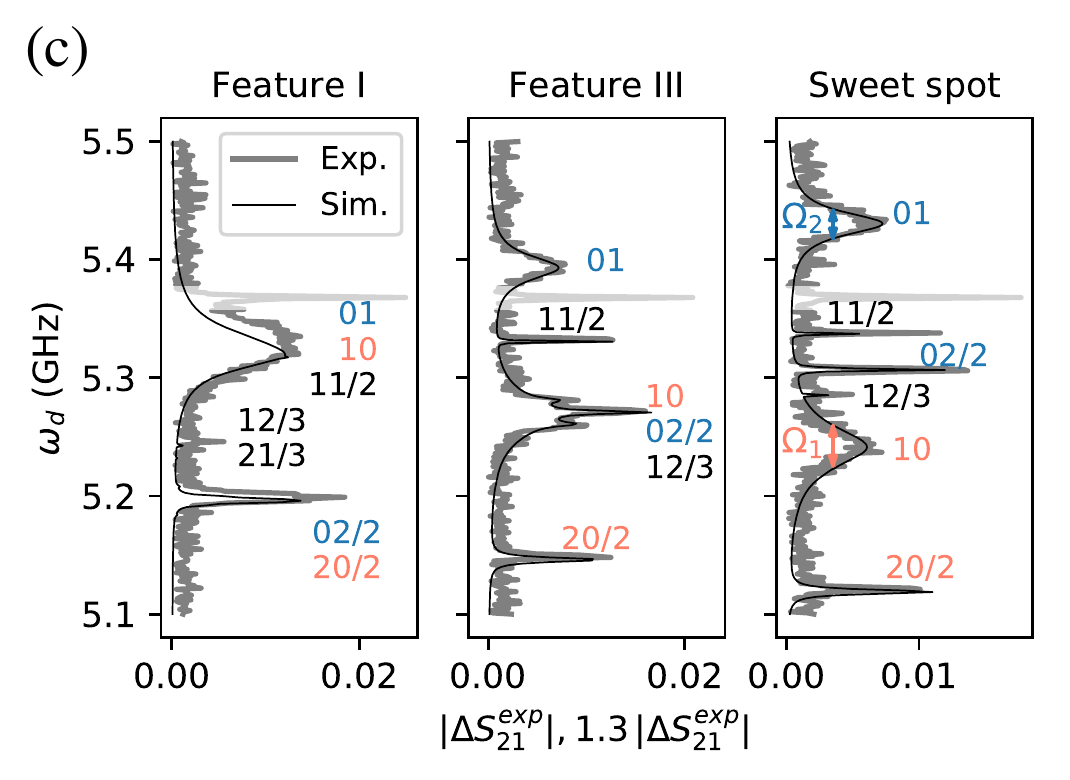}
	\includegraphics[width=.495\linewidth]{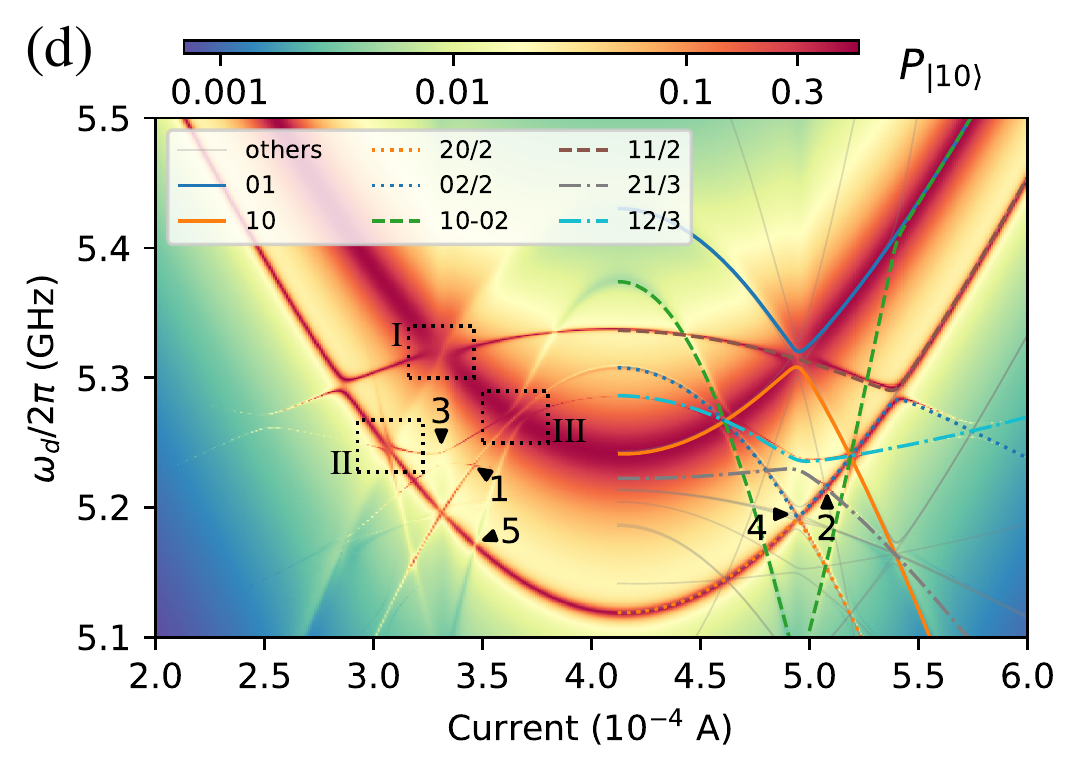}
	\caption{\textbf{(a)} 
	Measured spectroscopic data. Color shows the absolute deviation $|\Delta S_{21}| (I, \omega_d)$ of the complex transmission through the sample from its value in the lower left corner. 
	Two transmons are aligned as in 
	\autoref{fig:experiment}~(c) and form a 
	symmetric picture. Experimental data contain 
	an additional horizontal line from a 
	parasitic resonance interacting with the 
	readout resonator. Three effects
	not predicted by the unperturbed model are 
	shown in insets I, II and III; other 
	pronounced features reproduced by the 
	modeling are shown with Arabic markers (see 
	text for description). \textbf{(b)} Numerical 
	simulation reproducing experimental results 
	with labeled transitions (see Section 
	\ref{sec:analysis} for notation). The simulated $|\Delta S_{21}| (I, \omega_d)$ is multiplied by 1.3 to match with the experiment.
	\textbf{(c)} Slices of (a) and (b) at various 
	currents: 0.333 mA (feature I), 0.365 mA (feature III), 0.411 mA 
	(sweet spot); 
	experiment is gray, simulation is black. Individual Rabi 
	frequencies $\Omega_{1,2}$ may be extracted 
	as FWHM of the spectral lines 10 and 01, 
	respectively. \textbf{(d)} Simulated 
	steady-state population of the state $\ket{10}$ vs. $I$ and $\omega_d$. Lines 
	show various transition frequencies predicted by the 
	unperturbed Hamiltonian. In the legend, the main
	transitions are labeled near the sweet spot current; if 
	elsewhere any two lines form an avoided 
	crossing, the labels should be swapped after 
	the intersection. Under label ``others'' we put all the remaining secondary transitions (single- and multiphoton) between each pair levels of the system that fall into the relevant frequency range of the plot.}
	\label{fig:two-tone}
\end{figure*}

The experiment was conducted as described in Appendix 
\ref{sec:meas_setup}. We use high-power 
spectroscopy to probe transitions between the 
eigenstates of the SAM; the experimental data are shown in 
\autoref{fig:two-tone}~(a). As one can see, besides the 
fundamental lines that were shown in 
\autoref{fig:experiment}~(c), some 
new spectral ones are visible. Their frequency 
also depends on the applied current, and at 
several points they become resonant with each 
other. At three pairs of such resonant points, we 
observe distinct features shown in insets and marked 
with Roman numbers I, II, and III. Some secondary details are 
shown with Arabic-numbered markers. To avoid any possible confusion, we emphasize that by feature I we mean not the usual avoided crossing between $\omega_{1}(I)$ and $\omega_{2}(I)$, studied extensively in the past at lower powers. Instead, we call feature I its apparent disappearance and the noticeable change in the shape of the two-photon spectral line that would usually pass straight through it \cite{filipp2011multimode}. 

To check whether the standard theoretical model summarized in 
\autoref{eq:master} can reproduce the 
experimental spectrum, we have solved the master 
equation \autoref{eq:master} finding the steady 
states of the system $\hat \rho_{ss}(I, \omega_d)$ and the 
corresponding expected measurement outcomes 
$\Tr[\hat M \hat \rho_{ss}(I, \omega_d)]$ 
depending on the external coil current, and the excitation frequency $\omega_d$.
The results are shown in 
\autoref{fig:two-tone}~(b) where all the 
experimental details are immediately reproduced 
with just nine SAM states. We have solved \autoref{eq:master} in 
the rotating frame with RWA using 
\autoref{eq:steady} and in the lab frame (using 
the propagator approach \autoref{eq:propagator}) 
and did not find any noticeable difference in the 
results; though, the runtime of a $401\times401$ point simulation is 9 hours without RWA vs. 
3 minutes with RWA. The driving amplitudes $\Omega_{1,2}$ were $20$ and $10$ MHz, respectively. The parameters of the unperturbed Hamiltonian summarized in \autoref{tab:parameters} were established by fitting the system transition frequencies obtained from numerical diagonalization to the observed spectral lines in \autoref{fig:two-tone}~(a). The transmon interaction strength $J$ is usually determined from the size of the avoided crossing between $\omega_1$ and $\omega_2$. However, in \autoref{fig:two-tone}~(a) it is smaller than the linewidths and is not resolved due to the strong drive. By fitting separately measured data at lower power we have obtained $J = 8.69$ MHz. Alternatively, $J$ can be determined from the size of the  splitting located at 5.3 GHz and $2.9\cdot 10^{-4}$A (or $5.4 \cdot 10^{-4}$A). It is a well-known effect \cite{dicarlo2009demonstration} widely used for implementing cPhase gates on transmons; its size is $(2 \cdot J_{eff})/2$, where $J_{eff} = \sqrt{2} J$ is the corresponding matrix element of $\hat H_{int}$ and division by two is required as we observe an intersection between two-photon processes in our data. As can be seen from three 
slices of \autoref{fig:two-tone}~(a),~(b) shown in 
\autoref{fig:two-tone}~(c), the numerical results 
are in a good agreement with the experiment (the 
spurious resonance is softened).

When modeling the dispersive readout, we noticed that upon SAM excitation the resonance dip was reduced slightly in the experiment. Therefore, the actual observed response was higher than that predicted by only the dispersive shifts. Due to this fact, in \autoref{fig:two-tone}~(b),~(c) we had to artificially scale the theoretical data up approximately by 30\% to obtain the best match with the experiment. We believe that this problem is connected with the very low observed internal Q-factor of the readout resonator which was 100 times lower than the corresponding Q-factors of the test resonators and the transmons located on the same chip. We have observed this suppression of the readout resonator internal Q-factor for several similar devices, but so far we could not find an explanation for that. However, this problem does not affect our main results which only concern the behavior of the SAM.

Returning to the spectral data, we note that features I, II and III shown in the insets turn out to be 
impossible to explain using only the unperturbed 
Hamiltonian; we demonstrate this by numerical 
diagonalization of the unperturbed Hamiltonian used in the full simulation. In \autoref{fig:two-tone}~(d) we show all possible 
single and multiphoton transitions between all resulting pairs of the unperturbed eigenlevels that fall into the relevant frequency range. While correctly reproducing, for 
instance, the avoided crossing labeled as 
feature 3, the unperturbed model does not produce any transitions following the spectral lines observed in the Roman-numbered areas. For example, one can see that the unperturbed transition 11/2 passes straight in the middle of the avoided crossing between 01 and 10. The experiment and the full simulation demonstrate that the strong drive may change the observed shape of this line.

So far, we have established the fact that the 
same model SAM yields different spectra depending 
on the presence of the driving, and that the full 
model \autoref{eq:master} agrees well with 
experimental data. To understand the nature of features I, 
II and III, we describe \autoref{fig:two-tone} in 
more detail in a separate section below.

\begin{figure*}
	\includegraphics[width=.49\linewidth]{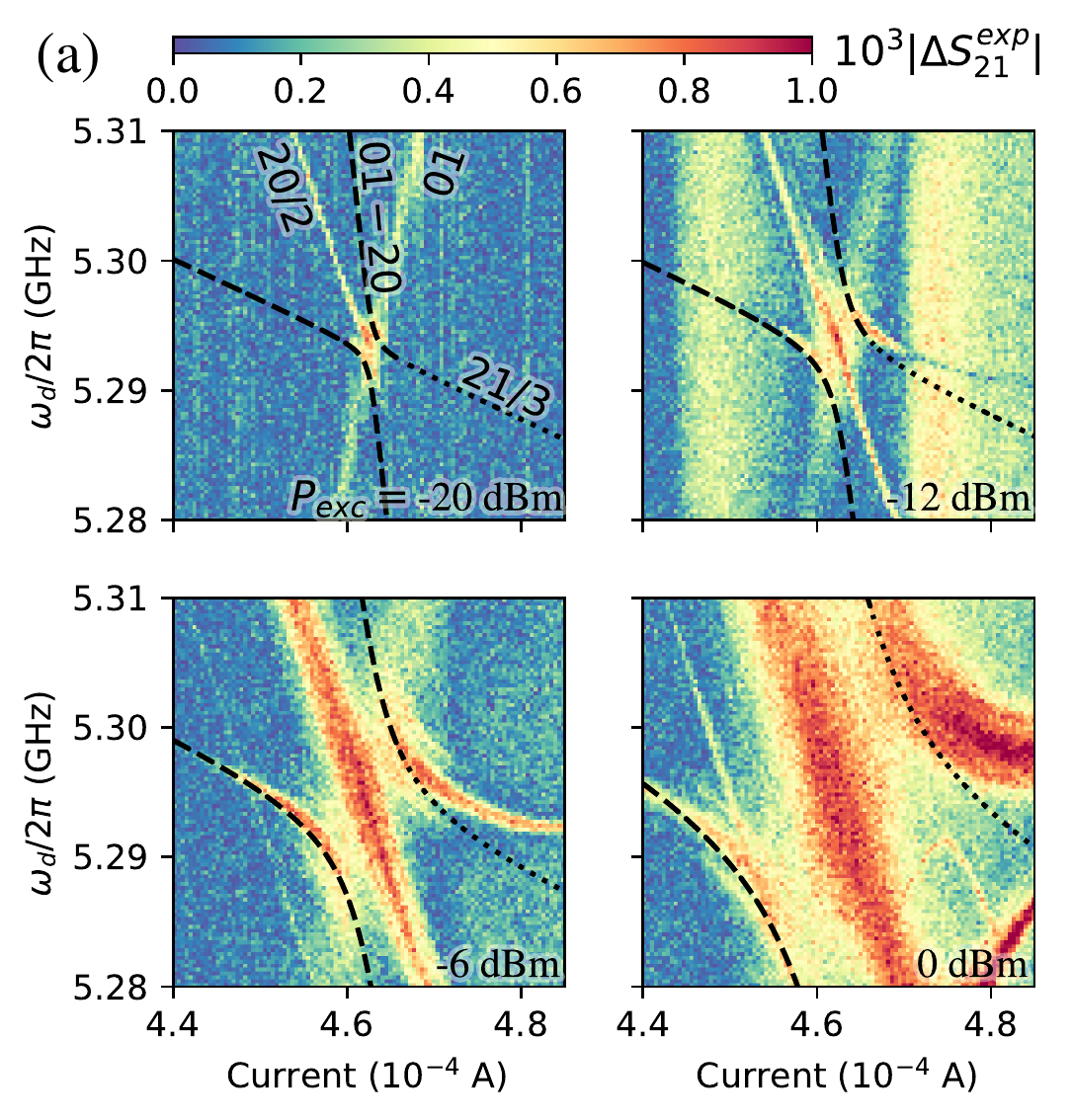}
	\includegraphics[width=.49\linewidth]{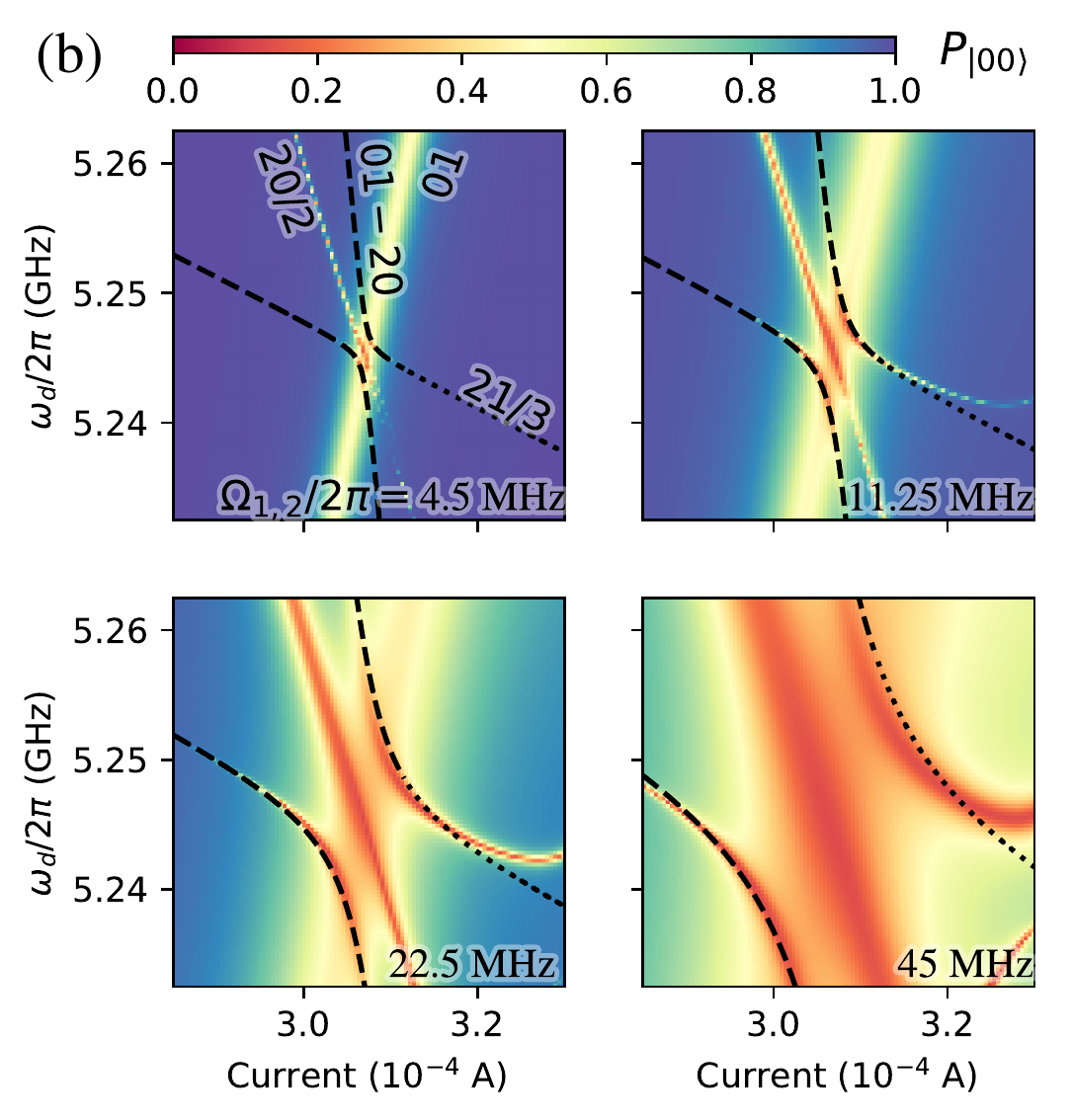}
	
	\includegraphics[width=.7\linewidth]{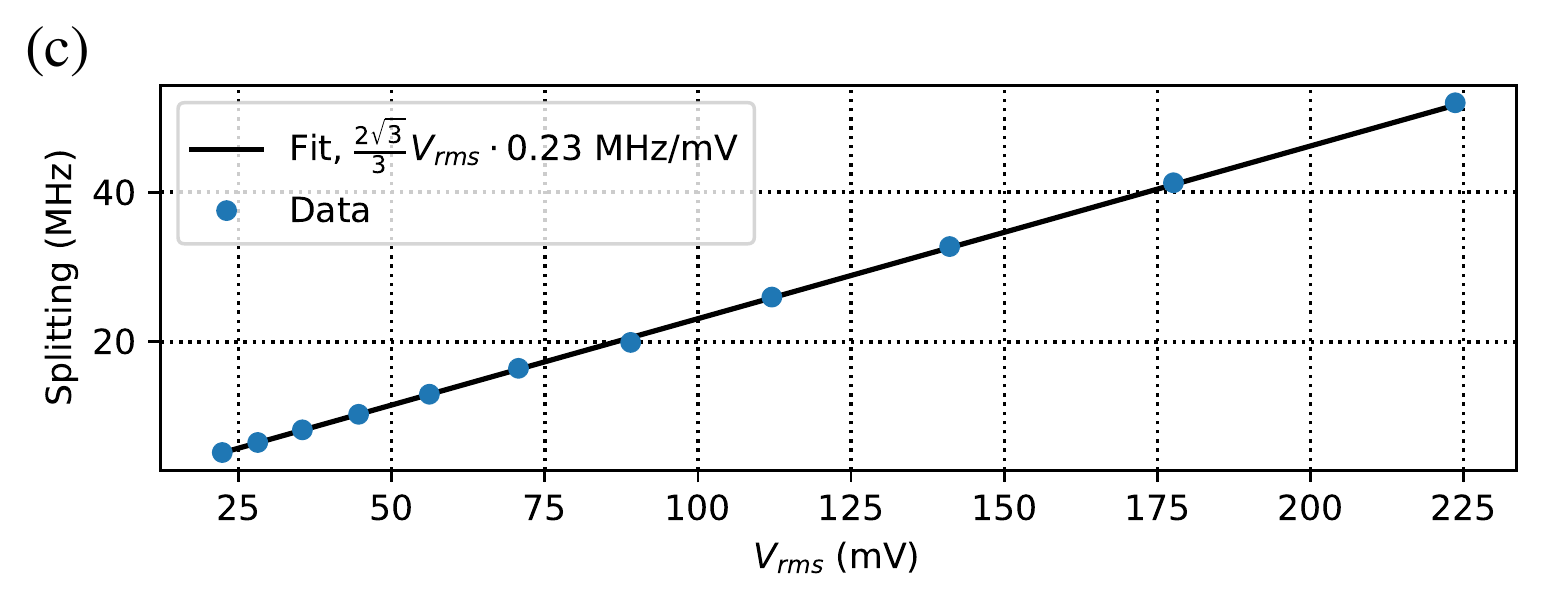}
	\caption{Power dependence of feature II: 
	experiment, simulation and analytical model. Dashed are the model curves turning to dotted when the 
	model is not expected to be valid (see 
	Sections \ref{sec:theory} and 
	\ref{sec:münchhausen}); model values for 
	$\Omega_{1,2}$ are the same for both (a) and 
	(b): $\Omega_{1,2}/2\pi=$ 4.5, 11.25, 22.5 
	and 45 MHz. \textbf{(a)} Experiment (second 
	cooldown; the system parameters are slightly different to 
	those in \autoref{fig:experiment}). The power 
	of the microwave source is increased from -20 
	to 0 dBm, and the corresponding growth of the 
	splitting and the widths of the spectral 
	lines is observed. \textbf{(b)} Simulation 
	with amplitudes of the driving $\Omega_{1,2}$ 
	equal to the model values; other parameters 
	as in \autoref{fig:two-tone}. Note that now 
	colors show the steady-state population of the ground 
	state. \textbf{(c)} Linear dependence of the 
	splitting size on the driving voltage of the 
	microwave source $V_{rms}$, ${\Omega_2}/{V_{rms}} = 2 \pi\cdot 0.23\ 
	{\text{MHz}}/{\text{mV}}$. From Section 
	\ref{sec:theory}, the splitting size is 
	$\frac{2\sqrt{3}}{3} \Omega_2$.}
	\label{fig:zoom}
\end{figure*}

\begin{figure*}
	\centering
	\includegraphics[width=\linewidth]{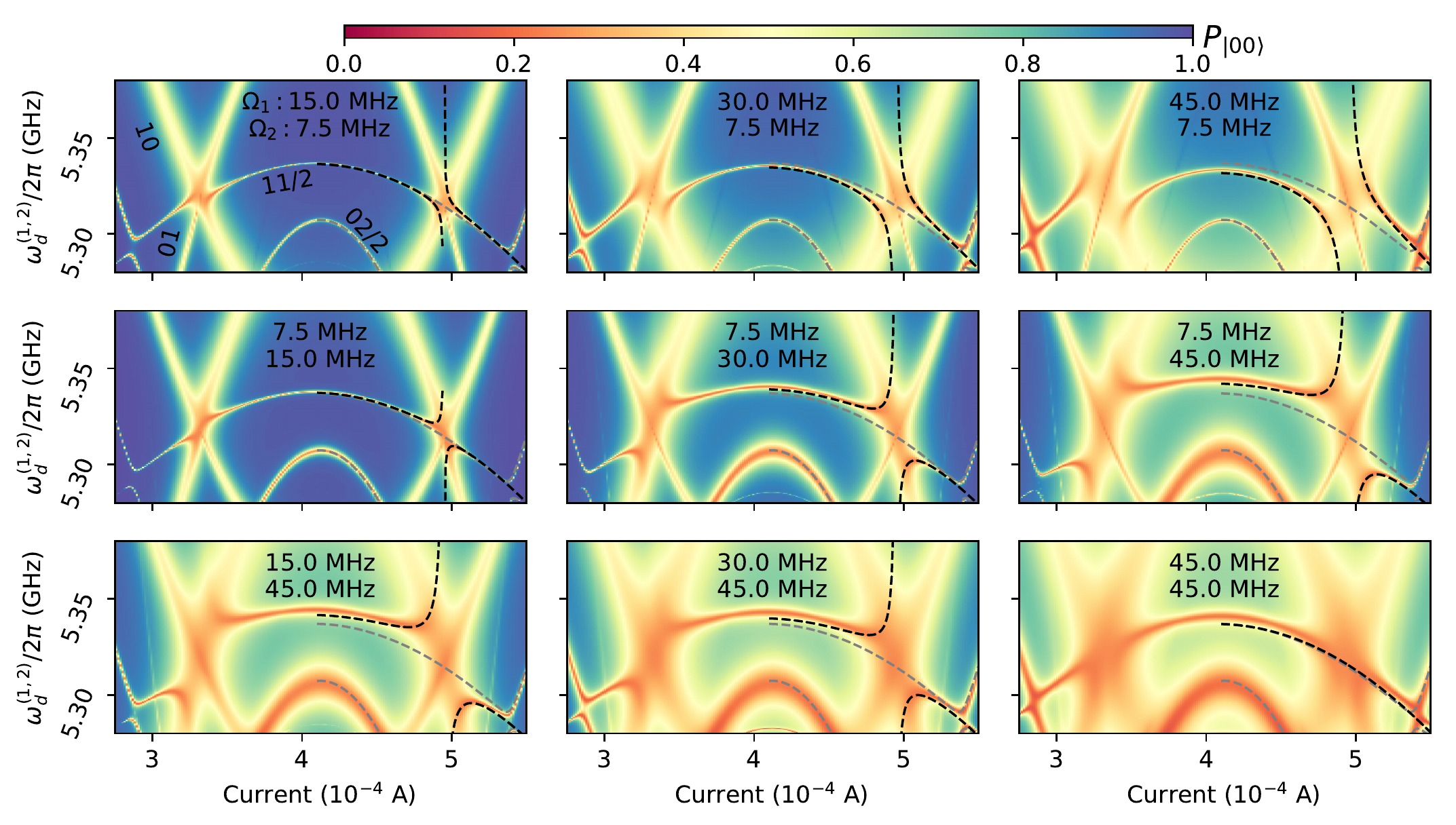}
	\caption{Simulated power dependence of feature I. With color we show the steady-state population of the ground 
	state when the transmons are driven at the 
	same frequency but at different amplitudes 
	$\Omega_{1,2}$. In the top row (middle row), 
	$\Omega_{1(2)}$ is increased while 
	$\Omega_{2(1)} = \text{const},\ \Omega_{2(1)} < 
	\Omega_{1(2)}$; two topologically different 
	types of splittings arise depending on 
	which transmon is driven stronger. In the 
	bottom row, we show how the splittings vanish 
	when the weaker drive is increased to match 
	with the stronger one. Gray dashed lines show 
	the unperturbed solution, in black are 
	the model curves based on Sections 
	\ref{sec:theory} and \ref{sec:münchhausen}.}
	\label{fig:difdrive}
\end{figure*}

\subsection{Analysis of the spectra} 
\label{sec:analysis}

\paragraph{Identification of the spectral lines.} 
Though it is not possible to describe all of the 
experimental features with the unperturbed model, 
we can still use it to identify most of the 
observed transitions as follows. By numerical diagonalization, we calculate the 
unperturbed frequencies of the undriven SAM as 
$\omega_{mn} = (E_n - E_m)/h$ where $n>m,\  
n,m=0..8$. In \autoref{fig:two-tone}~(d), we 
place them over the heatmap data optionally dividing 
$\omega_{mn}$ by some integer $n$ for $n$-photon processes. 
Since the number of SAM states is finite, we can 
quickly find all matches between the heatmap and the calculated eigenlevels. 

Next, at the sweet spot, we label the found lines 
as $ij/n$. Here, $i,j$ denote the occupation numbers of the transmons 1 and 2, respectively, in the destination 
state $\ket{ij}$ of the transition; $/n$ is optional for 
an $n$-photon process. Such labelling is meaningful because at 
the sweet spot the transmons are detuned from 
each other, and thus SAM eigenstates are nearly 
factorized. Using this 
notation, we label lines in 
\autoref{fig:two-tone}~(b),(c), as well. The notation is only valid 
 near the sweet spot current before any two lines 
form an avoided crossing; when they do, their
names should be swapped.
 
As one can see, various multiphoton transitions besides 
the main lines at $\omega_{1,2}(I)/2\pi$ (or 
$01$ and $10$ lines) are visible. The ${02/2}$ and ${20/2}$ 
are very commonly found for transmons and lie 
$|\alpha_{1,2}|/2$ lower than the main lines 
$\omega_{1,2}(I)$. Another two-photon process is the $11/2$ line when two SAAs are excited 
simultaneously. This process is used, for 
example, for the bSWAP 
gate \cite{poletto2012entanglement}; in this work, 
it is taking part in the formation of feature I. A 
three-photon process $12/3$ is also clearly 
visible just below the $02/2$ line. As we will 
see, processes $12/3$~and~$21/3$ are involved in forming features III and II, respectively. Notably, transitions $11/2$, ${21/3}$, and 
${12/3}$ are forbidden when there is no 
interaction between transmons ($J=0$). Therefore, 
it is expected that all Roman-numbered features 
should only appear with non-vanishing $\hat 
H_{int}$.

There are also some avoided crossings predicted by the unperturbed model. For a two-transmon SAM they occur ar the intersection between lines 
${01}$, ${10}$ or $ {11} $, $ {02} $ ($ {20} $) 
visible clearly in \autoref{fig:two-tone}~(d). 
Additionally, marked as feature 3 in 
\autoref{fig:two-tone}, we see an avoided 
crossing between ${12}/3,\ {21}/3$ at the same 
current where ${01}$, ${10}$ intersect.

\paragraph{Analysing features I, II and III.} 
First, we have reproduced the avoided crossing of feature III in an additional numerical simulation taking only 2 levels for the transmon 1 and three levels for 
the transmon 2. Upon this, it has become clear 
that features II and III are actually of 
the same nature and differ only by the ordering 
of the transmons: for II, they appear when the 
transition ${01}$ intersects ${20/2}$, and for III when ${10}$ intersects ${02/2}$; 
one can see this clearly in 
\autoref{fig:two-tone}~(b).

From \autoref{fig:two-tone}~(d) we conclude that 
avoided crossing in III is between two 
transitions: ${12/3}$ (three-photon transition $\ket{00}\rightarrow\ket{12}$) and ${10} - {02}$ 
(single-photon $\ket{10}\rightarrow\ket{02}$) which are of the 
same frequency when $\omega_1 = 
\omega_2+\alpha_2/2$. The latter process is depopulating $10$ and it is better discernible in \autoref{fig:two-tone}~(d) 
than in \autoref{fig:two-tone}~(a), (b). For II, the 
opposite is true: ${21/3}$ and ${01} - {20}$ are 
crossing when $\omega_2 = \omega_1+\alpha_1/2$. 
From additional measurements and simulations, we 
find that the splitting depends on the driving 
power; the experimental and simulated results for 
feature II are shown in color in \autoref{fig:zoom}~(a), (b), respectively. As one can see, the growth of the splitting with 
increasing power is linear: it is roughly equal 
to the FWHM of the ${01}$ spectral line. To fully 
quantify the shape of this splitting, in Section 
\ref{sec:theory} we derive analytical expressions 
for the dashed curves fitting the spectral lines. In \autoref{fig:zoom}~(c), blue points, we present splitting sizes extracted by fitting that analytical model to the data as in \autoref{fig:zoom}~(a) for various power values of the microwave generator connected to the excitation waveguide. As one can see from the linear approximation of the points, the splitting indeed is simply proportional to the $V_{rms}$ of the signal. From the model, we expect that the minimal distance between the branches of the avoided crossing is equal to $\frac{2\sqrt{3}}{3} \Omega_2$; using this relation, we extract the proportionality coefficient between $\Omega_2$ and $V_{rms}$ of around $2\pi \cdot 0.23$ MHz/mV.

In \autoref{fig:difdrive}, we demonstrate a simulated power dependence of feature I: increasing the drive amplitude on one of the 
transmons while keeping the other one small and 
constant again result in a certain splitting of the line 11/2. We show only the calculation as long as an experiment is not possible with our sample since we only have a single excitation line and there is 
no way for us to control the driving amplitudes 
$\Omega_{1,2}$ independently. We note that two 
qualitatively different patterns arise depending 
on which of the transmons is driven stronger than 
the other. From this and from the shape of the 
splitting in feature I of 
\autoref{fig:two-tone}~(a), we can infer that 
$\Omega_1/\Omega_2 \approx 2$ there (also 
consistent with the 01, 10 linewidths in 
\autoref{fig:two-tone}~(c)). If in contrast both transmons are driven with equal amplitudes, the splitting of 11/2 vanishes. As can be seen from the black dashed lines in \autoref{fig:difdrive}, all these cases are explained well by our analytical model described in detail in Sections \ref{sec:theory} and \ref{sec:münchhausen}.

From all presented observations, we conclude that effects I -- III 
are caused by light dressing. In case III, the 
first transmon is dressed by a strong resonant 
field; in case II, the second one; finally, in 
case I, both transmons may be dressed at the same 
time. We will discuss these effects in greater 
detail in Sections \ref{sec:theory} and \ref{sec:münchhausen}.

\paragraph{Secondary features.} Using 
\autoref{fig:two-tone}~(d), we can get an insight 
into the features 1 -- 5 as well. 

Feature 1 is a small avoided crossing between 
${02/2}$ and ${21/3}$. It is missing in the 
unperturbed solution and thus is caused by the 
light dressing just as I -- III. Feature 2 is its 
twin: ${12/3}$ and ${20/2}$ intersect there but 
the anticrossing is smaller due to the asymmetry 
of the driving strengths and is not resolved. 

Feature 3 is a large avoided crossing between 
three-photon processes ${12/3}$ and ${21/3}$. It 
is predicted by the unperturbed model, and direct 
diagonalization yields the splitting of 
$\frac{4}{3}J$. A remarkable detail here is that 
the dim lower branch implies the presence of a 
dark state with respect to the driving operator 
in the third order. 

Feature 4 is also explained by the unperturbed 
model and is caused by several spectral lines and 
a pair of avoided crossings near a single point 
(dotted lines in \autoref{fig:two-tone}~(d)). It 
appears at the point where ${02/2}$ intersects 
${20/2}$ and is just barely visible in the 
experimental data because of the noise. 

Feature 5 is located at the intersection between ${20/2}$ 
and ${10} - {02}$, and can be 
found in the experimental data, too. 

In conclusion to this section, we note that when the coupling is turned off ($J=0$) in the simulation, the system does not demonstrate any of the described details. This means that all these effects can only be attributed to the SAM as a whole. 

\subsection{Explaining Roman-numbered effects}\label{sec:theory}

Since we had already connected the additional 
spectral features with light dressing, it is 
natural to expect an Autler-Townes-like effect to 
be at the root of the additional spectral lines. 
For a three-level system, the standard A-T effect 
is revised in Appendix \ref{sec:3-level-at}. 
However, in our case the level structure and the 
effect itself are more complicated. 

First of all, since during the spectroscopy we 
apply only a single microwave tone ($\omega_d  = \omega_d^{(1)} = 
\omega_d^{(2)}$), it has to be 
simultaneously the coupler and the probe
in the terminology of the standard A-T effect; moreover, the 
probe must be much weaker than the coupler. 
It turns out that these conditions become satisfied around feature II (III) when 
$\omega_{2(1)} = 
\omega_{1(2)}+\alpha_{1(2)}/2$ where we 
can simultaneously excite transitions $01 (10)$ 
and $20/2 (02/2)$. Since the two-photon Rabi 
frequency is much smaller than the single-photon one, it is natural to view the two-photon 
excitation as the probing process which does not 
affect the level structure. In contrast, the
single-photon excitation is strong enough to 
dress the system. In other words, for the feature III, the 
A-T coupling operator is $\hat H_{d}^{(1)}$, 
and the probing operator is $\hat H_{d}^{(2)}$. However, their 
separation now is not in frequency, but 
in the Hilbert subspaces they act upon and in the number of participating photons.

For the feature I, the simultaneously 
excited transitions are $01,\ 10$ and the 
two-photon $11/2$. In this case, the weak two-photon process is probing transitions in the doubly-dressed SAM. We find that if SAAs are dressed equally, $11/2$ does not split, and for it to be distinguishable it is required that 
$\Omega_{1(2)} \gg \Omega_{2(1)}$. 

Below, we give a detailed explanation of features II and III, and finally, I.

\begin{figure}
	\centering
	\includegraphics[width=\linewidth]{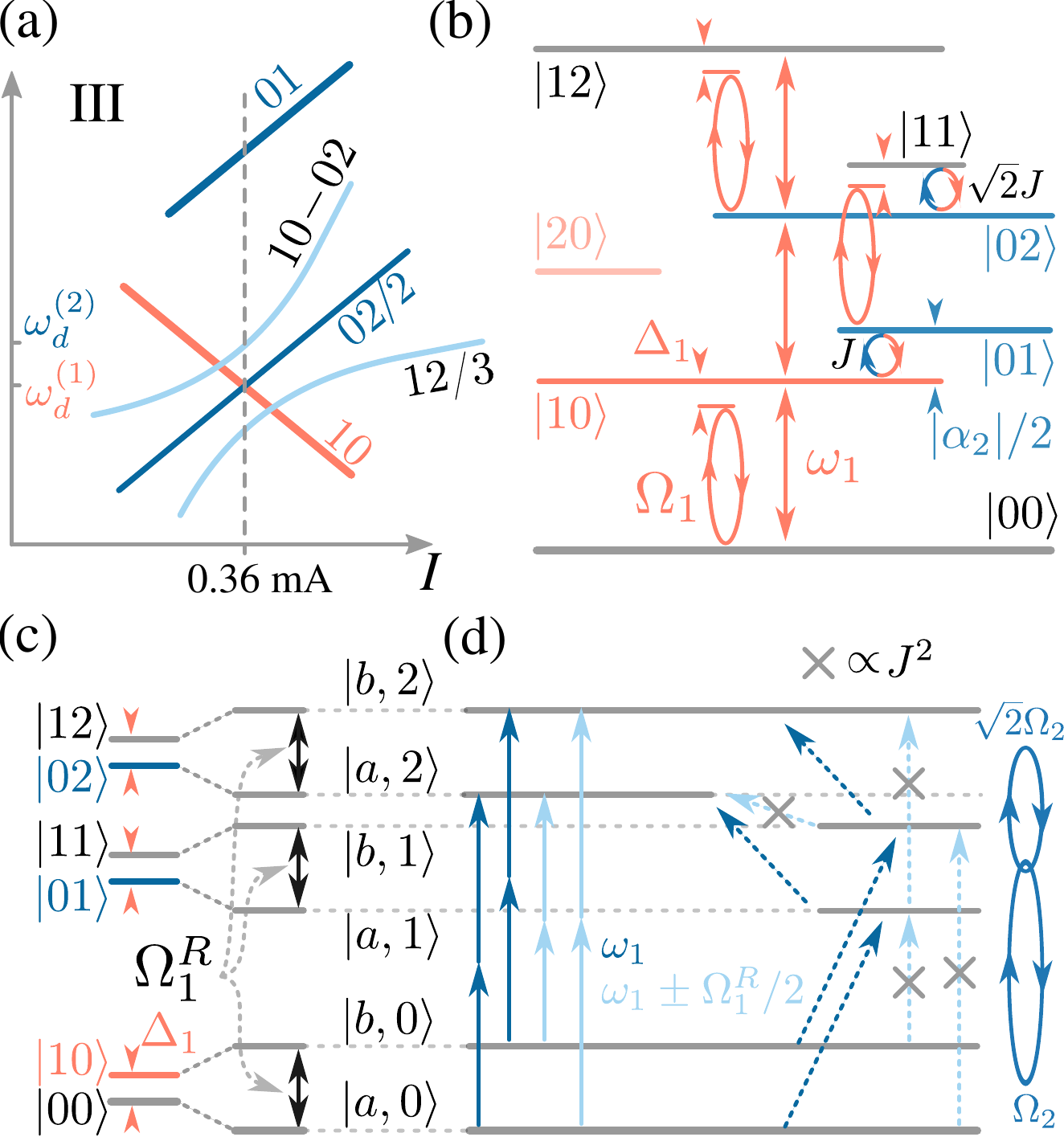}  
	\caption{\textbf{(a)} 
	Schematic of the transition frequencies near 
	III (not to scale). Resonant point is marked by a dashed gray line. Here we assume 
	that each transmon is driven at its own 
	frequency. \textbf{(b)} System energy levels at the resonant point; the first transmon 
	drive at a small detuning 
	$\Delta_1$ from $\omega_1$ is shown as orange ellipses. The second 
	transmon driving is not shown here. Action of 
	$\hat H_{int}$ in RWA is depicted as 
	orange-blue circles. \textbf{(c)} In the 
	frame rotating with $\hat 
	H_{d}^{(1)}$, states $\ket{0j},\ket{1j}$ 
	become nearly degenerate ($\omega_1 
	\rightarrow \Delta_1$). Dressing increases this splitting to 
	$\Omega_1^R$. \textbf{(d)} Transitions in the 
	dressed system induced by $\hat 
	H_{d}^{(2)}$ (coupled level subspaces are shown with blue ellipses). In 
	the left part of the panel, all possible 
	two-photon transitions near $\omega_1$ are 
	depicted: the blue transitions are not 
	shifted in frequency, the light blue ones are shifted 
	by $\pm\Omega^R_1/2$. In the right 
	part, some of the contributing trajectories are 
	depicted; gray crosses show transitions that 
	are forbidden without the coupling $J$.}
	\label{fig:main_scheme}
\end{figure}

\paragraph{Features II and III.}  

In \autoref{fig:main_scheme}, we fully identify the transitions between dressed states that
are involved in the A-T-like effect for feature III. 
For simplicity, we will temporarily assume that the transmons are driven at different frequencies $\omega_d^{(1)}$ and $\omega_d^{(2)}$
(this assumption will be lifted in the following 
section). For convenience, in 
\autoref{fig:main_scheme}~(a) we schematically 
reproduce the transitions participating near III where the 
first transmon is below the second one, and the current of the resonance point where $\omega_{1} = 
\omega_{2}+\alpha_{2}/2$ is 
shown with a dashed gray line. Next, in
\autoref{fig:main_scheme}~(b) we plot the 
level structure of the SAM at the resonance point. The single-photon drive by 
$\hat H_d^{(1)}$ detuned by 
$\Delta_1$ from $\omega_1$ is shown with orange ellipses. The
much weaker two-photon drive by $\hat H_d^{(2)}$ is not shown yet because it does not alter the structure of the energy levels. Since we know from the numerical simulations that the third level of the first transmon is not 
necessary to observe the splitting, $\ket{20}$ is 
shown transparent, and states $\ket{21},\ 
\ket{22}$ are not shown.

The next step is to view the system in the frame 
rotating with the first transmon and then move to 
the  dressed picture similarly to Appendix 
\ref{sec:3-level-at}. Now, the first transmon 
splitting equals $\Omega_{1}^R = 
\sqrt{\Omega_1^2+\Delta_1^2}$, and its new 
eigenstates (dressed states) are denoted as 
$\ket{a}$ and $\ket{b}$. Meanwhile, the second 
transmon subspace is not altered. The corresponding level system in the rotating frame before and after modification by the drive is shown in 
\autoref{fig:main_scheme}~(c).

Finally, in \autoref{fig:main_scheme}~(d) we demonstrate possible two-photon transitions between the dressed states induced by the second transmon driving $\hat H_{d}^{(2)}$. In the left part of the panel, one can find the unmodified two-photon transition $02/2$ at $\omega_1$ and two sidebands at $\omega_1 \pm \Omega_1^R/2$. This picture finally explains the observed triplet transition of feature III. The right part describes the mechanism of these two-photon processes through virtual excitations of the intermediate states. From here it becomes obvious that without the transmon-transmon interaction the sideband transitions are forbidden due the selection rule: $\bra{a,j}\hat{\mathbbm{1}}\otimes \hat H_d^{(2)} \ket{b, j+1} = 0$ since $\braket{a}{b} = 0$. However, we will show that they become allowed in the second order in $J$ when the coupling is turned on.

Now, we will repeat this reasoning 
quantitatively. We start from 
the initial Hamiltonian, \autoref{Hsystem}. To 
move to the rotating frame with 
\autoref{eq:rotation} and apply the RWA, we use 
the following operator:
\begin{equation}
R = \exp[-it \omega_d^{(1)} 
(b^{\dagger}b+c^{\dagger}c)].
\end{equation}  
Note that now we rotate both transmon subspaces simultaneously in contrast to what is shown in \autoref{fig:main_scheme} because below it will be convenient to have time-independent $\hat H_{int}$. Now,
\begin{equation}
\begin{aligned}
\omega_{1,2} &\rightarrow \Delta_{1,2} = \omega_{1,2} - \omega_d^{(1)},\\
\hat H_{int} &\rightarrow  \hbar J \left[\hat \sigma_+ \hat c + \hat \sigma_-\hat c^\dag \right],\\
\hat H_{d}^{(1)} &\rightarrow \frac{\hbar \Omega_1}{2} \hat \sigma_x,\ 
\hat H_{d}^{(2)} \rightarrow \frac{\hbar \Omega_2}{2}(\hat c e^{i\delta t}  + \hat c^\dag e^{-i\delta t}),
\end{aligned}
\end{equation}
where $\delta = \omega_{d}^{(2)} - \omega_{d}^{(1)}$.

Since $\hat H_{d}^{(1)}$ is now time-independent, we can move to the dressed basis by applying a transformation $\hat S$ which diagonalizes the first transmon. After that, the Hamiltonian may be split into three parts
\begin{equation}
\begin{aligned}
\hat H^{D}/\hbar &= \left[\frac{\Delta_1}{2} 
\hat{\mathbbm{1}}-\frac{\Omega_1^R}{2}\hat 
\sigma_z\right] + \Delta_2 \hat b^{\dagger}\hat b 
+ \frac{1}{2}\alpha_2 \hat b^{\dagger}\hat b(\hat 
b^{\dagger}\hat b-1),\\
\hat V_J &= \hat S^\dag \hat H_{int}\hat S,\\
\hat V_t (t)&= \hat S^\dag \hat H_{d}^{(2)} \hat S \equiv \hat H_{d}^{(2)}.
\end{aligned}
\end{equation}
The first part is diagonal, and the remaining two 
will be treated as perturbations. To simplify 
further calculations, we consider the point where 
$\Delta_1 = 0$ and $\Delta_2 = |\alpha_2|/2$. In 
these conditions, $H^D$ becomes degenerate as can 
be seen in \autoref{eq:ham_matrix} of Appendix 
\ref{sec:dpt}. Using the degenerate perturbation 
theory for $\hat V_J$ summarized therein, we find 
the first-order corrected wave functions of $\hat 
H^D + \hat V_J$, labeled $\ket{k},\ k=1..6$. We find the mean relative element-wise error of around 3\% between numerically obtained eigenvectors and perturbative ones for our experimental parameters.

The time-dependent perturbation theory that we use to calculate the transition rates of the two-photon processes stimulated by $\hat V_t$ is reviewed in Appendix \ref{sec:2pp}. Using the corrected eigenstates $\ket{k}$ and neglecting small terms, we obtain the following expressions for the transition rates per unit time \cite{faisal2013theory}:
\begin{equation}
\begin{aligned}
R^{(2)}_{2\rightarrow 5} &\approx \pi\Omega_2^4 
\frac{16 J^4 \left(3 \alpha^2 + 
\text{$7\Omega_1^2$}\right)^2}{\alpha 
^{10}},\\
R^{(2)}_{1\rightarrow 6} &\approx \pi\Omega_2^4 
\frac{4 J^4 \left(5 \alpha - \text{$3\Omega_1
		$}\right)^2}{\alpha ^{8}},
\end{aligned}\label{eq:rates}
\end{equation}
when $|\alpha_2| \gg \Omega_1, \Omega_2, J$ which is a good approximation for our setup. Taylor expansion here leads to errors of less than 0.5\%.  From \autoref{eq:rates} 
follows that the sideband transitions are 
prohibited without the interaction in the SAM 
$(J=0)$, and the extra avoided crossings will not 
be observed.

The above reasoning can be repeated for feature II without modification because it is only different from III in the ordering of the transmons.

\begin{figure}
	\includegraphics[width=.9\linewidth]{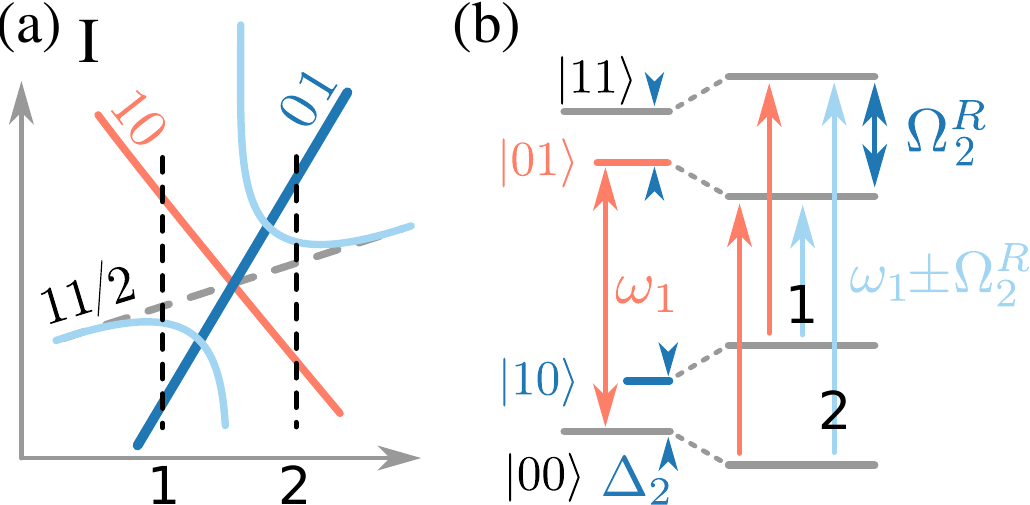}
	\caption{Feature I for the case of $\Omega_2 \gg \Omega_1$. (a) Schematic of the spectral lines near I (the avoided crossing between 01 and 10 is not shown). Compare to \autoref{fig:difdrive}, middle row. (b) Transitions between dressed states in the frame rotating with the second transmon. Two sideband transitions (light blue) appear at frequencies $\omega_1\pm\Omega_2^R$; however, only one of them may be observed at each side of the intersection for a monochromatic signal (see Section \ref{sec:münchhausen}).}
	\label{fig:featureI}
\end{figure}

\paragraph{Feature I.} Finally, we discuss the 
remaining Roman-numbered effect. In the case when 
$\Omega_2 \gg \Omega_1$ (the second transmon is 
dressed) the splitting has the shape shown in \autoref{fig:featureI}~(a). For the opposite case 
($\Omega_1 \gg \Omega_2$), the logic is similar. 

As one can see from \autoref{fig:difdrive}, near the 01, 10 intersection only the two-photon transition $11/2$ is affected and deviates from the unperturbed spectrum while the spectral lines 01, 10 do not shift (even though the avoided crossing between them vanishes). Similarly to II and III, in the frame rotating with the second transmon it turns into two different single-photon transitions for $\hat H_d^{(1)}$ located at $\omega_1 \pm \Omega_2^R$, see \autoref{fig:featureI}~(b). We will show below that in the experiment with a single excitation frequency, it is not possible to observe both transitions simultaneously. This is also clear from \autoref{fig:difdrive} where only one spectral line is present to the left and one to the right from the 01, 10 intersection.

\subsection{Self-consistent equations for the I, II, III}
\label{sec:münchhausen}

In the previous section, we have assumed that the transmons are driven independently at two different frequencies. In this section, we will discuss the realistic case when only a single frequency $\omega_d$ is sent at the SAM.
\paragraph{Features II, III.} We start again with feature III. As before, from 
\autoref{fig:main_scheme}~(d) we know that at the intersection the sideband 
transitions are formed by the two two-photon processes 
with $\hat H_d^{(2)}$. But, on the other hand, from 
the unperturbed solution we know that in the laboratory 
frame these two sideband transitions beyond the avoided 
crossing in III become $10 - 02$ (one-photon) and $12/3$ 
(three-photon). This means that there should be a 
smooth transformation between these one-, two- and three-photon 
processes when the system approaches the resonant 
point where $\alpha_2 + 2 \omega_{2} = 2\omega_1$. Let us consider hypothetical two-photon transitions $(10 
- 02)/2$ and $12/2$. In the frame rotating with 
the first transmon like in 
\autoref{fig:main_scheme}~(c) when $\Omega_1$ = 
0, their 
frequencies are
\begin{align}
\omega_{(10-02)/2} &= (\alpha_2 + 2 \omega_{2} - \Delta_1)/2,
\label{eq:two-photon_lab1002}\\
 \omega_{12/2} &= (\alpha_2 + 2 \omega_{2} + \Delta_1)/2.
\label{eq:two-photon_lab12}
\end{align}
When the first transmon becomes dressed by $\Omega_1 \neq 0$, its splitting $\Delta_1$ changes to $\Omega^R_1 =\sqrt{\Omega_{1}^2 + \left(\omega_{1} - \omega_{d}\right)^{2}}$. Substituting $\Omega^R_1$ instead of $\Delta_1$ into the equations above we note that $\omega_{(10-02)/2}$ and $\omega_{12/2}$ are now exactly equal to the two-photon sideband frequencies $\omega_1 \pm \Omega_1^R/2$ established in the previous section. Therefore, it is logical to use them to model the splitting behavior beyond the resonant point.

Since we are dressing the energy levels by $\hat H_d^{(1)}$ and probing two-photon transitions between them with $\hat H_d^{(2)}$ both at $\omega_d$, self-consistent equations have to be solved to find at which detuning the sideband spectral lines will appear:
\begin{equation}
\begin{aligned}
\omega_{(10-02)/2} &= \omega_d,\\
\omega_{12/2} &= \omega_d,
\end{aligned}
\label{eq:two-photon}
\end{equation}
where the left hand side for each equation is calculated according to \autoref{eq:two-photon_lab1002} and  \autoref{eq:two-photon_lab12}, respectively.

First, we analyse the solution of \autoref{eq:two-photon} in the case of no driving, $\Omega_1 = 0$. Both equations yield two identical answers for $\omega_d$ due to the fact that  $\Omega_1^R = |\omega_1 - \omega_d|$ this case. One can identify them as the frequencies of the three-photon $12/3$ and the single-photon $10-02$ transitions in the laboratory frame:
\begin{equation}
\omega_d^{(III, 0)} = \begin{cases} (\alpha_2 + 2\omega_2 + \omega_1)/3, \\ \alpha_2 + 2 \omega_{2} - \omega_{1}.\end{cases}
\end{equation}
Thus is a correct solution for the edge case since with no dressing, these transitions just intersect without an avoided crossing.

Next, for the general case of non-zero driving, $\Omega_1 \neq 0$, we obtain again two identical pairs of solutions:
\begin{equation}
\begin{aligned}
\omega_d^{(III, \pm)} = &\frac{2 \alpha_2}{3} + \frac{4 \omega_{2}}{3} - \frac{\omega_{1}}{3} \\ &\pm \frac{\sqrt{3 \Omega_{1}^{2} + \left( \alpha_2 + 2 (\omega_{2} - \omega_{1})\right)^{2}}}{3}.
\end{aligned}
\label{eq:splitting_model}
\end{equation}
These new curves do not intersect at any point, correctly reproducing the behavior observed in the experiment and in the numerical data and forming two branches of an anticrossing. The minimal splitting is 
found at the resonant point and equals $\frac{2 
\sqrt{3}}{{3}} \Omega_1$. A similar result can be produced for the feature II just by 
swapping the transmons; in that case, the 
splitting will be $\frac{2 \sqrt{3}}{{3}} 
\Omega_2$.

We find excellent agreement between the self-consistent solution \autoref{eq:splitting_model} and 
both the experimental and simulated data as can 
be seen in \autoref{fig:zoom}. To plot the model 
curves, we have approximated the 
$\omega_{1,2}(\Phi_e)$ transitions by  
linear functions and substituted them in 
\autoref{eq:splitting_model}. $\alpha_1$ is known 
from spectroscopy leaving only $\Omega_2$ to be found by fitting. We have repeated this 
approximation for a range of microwave powers to 
confirm the linear dependence of the splitting on the 
excitation amplitude (see \autoref{fig:zoom}~(c)).

The deviation of the upper branch from the data (see \autoref{fig:zoom}, dotted lines) is expected and caused by the avoided crossing marked as the secondary feature 3; it could be taken into account by modifying correspondingly \autoref{eq:two-photon_lab12}. The small discrepancy between the experimental and simulated data in the upper branch is caused by a slight elevation of the lower sweet spot of the first transmon moving the avoided crossing 3 closer to the feature II in the second cooldown.

\paragraph{Feature I.} Feature I may be explained using the same dressing model. Looking at \autoref{fig:featureI} for the case $\Omega_2 \gg \Omega_1$, we can write another pair of self-consistent equations:
\begin{equation}
\omega_{1} \pm \Omega_2^R = \omega_d^{(I, -)}.
\end{equation}
Substituting $\Omega_2^R = \sqrt{\Omega_2^2 + (\omega_2 - \omega_d^{(I, -)})^2}$, one can find that this system has a single solution
\begin{equation}
\omega_d^{(I, -)} = \frac{\omega_1 + \omega_2}{2} - \frac{ \Omega_{2}^{2}}{2 \left(\omega_{1} - \omega_{2}\right)},
\label{eq:topo_1}
\end{equation}
which turns into the frequency of $11/2$ in the laboratory frame when $\Omega_2$ is zero. When $\Omega_2$ is non-zero, the solution near the point $\omega_1 \approx \omega_2$ is a hyperbolic curve. However, in reality we do not observe the asymptotically vertical parts since the model becomes invalid when $|\omega_1 - \omega_2| \leq J$ due to the finite coupling between the transmons, and thus the avoided crossing between 01 and 10. Since $\omega_1$ and $\omega_2$ never reach each other, the denominator in \eqref{eq:topo_1} is limited from below and there is no actual divergence.

When $\Omega_1 \gg \Omega_2$, besides replacing $\Omega_2 \rightarrow \Omega_1$, 
one have to change the sign before the hyperbolic part to obtain the corresponding solution:
\begin{equation}
\omega_d^{(I,+)} = \frac{\omega_1 + \omega_2}{2} + \frac{ \Omega_{1}^{2}}{2 \left(\omega_{1} - \omega_{2}\right)},
\label{eq:topo_1_inv}
\end{equation}
yielding the reversed shape of the splitting.

Continuing this logic, we can write down the resonance condition in the doubly-rotating frame when both transmons are dressed simultaneously:
\begin{equation}
\Omega_1^R \pm \Omega_2^R = 0.
\label{eq:zero-photon}
\end{equation}
Solving it for $\omega_d$, we obtain the 
generalization of \autoref{eq:topo_1} and 
\autoref{eq:topo_1_inv}:
\begin{equation}
\omega_d^I = \frac{\omega_{1} + \omega_{2}}{2} + \frac{\Omega_{1}^{2} - \Omega_{2}^{2}}{ 2\left(\omega_{1} - \omega_{2}\right)}
\label{eq:topo_comm}
\end{equation}

\autoref{eq:topo_comm} is used to plot the black 
dashed curves in \autoref{fig:difdrive} taking 
the $\Omega_{1,2}$ values from the known simulation 
parameters shown therein. The frequencies 
$\omega_{1,2}$ are extracted from the fits of the spectral lines 01 and 10. We again find good agreement 
between the model and the simulated data.

\section{Discussion}

We have performed spectroscopic measurements of an 
isolated diatomic superconducting artificial 
molecule in the regime of strong interaction 
with classical light. Using joint dispersive 
readout to directly access the population of the SAM 
eigenstates, we have located several anomalies in the spectrum not explained 
by the unperturbed model of the system. Finding numerically the steady state of the SAM interacting with the classical drive, we have 
reproduced the experimentally discovered effects 
and attributed them to an altered version of the 
well-known Autler-Townes effect.

In contrast to the standard A-T effect, where there are two laser beams of different frequencies and powers, in our case, there is only a single tone interacting with the system. Therefore, 
it has to be both the coupler tone and the probe 
tone at the same time. However, since there are 
two components in the driving operator (one for 
each transmon), the separation between the 
coupler and the probe is still possible and occurs both in the Hilbert space and in the number of photons involved 
rather than in frequency and amplitude. The 
effectively weak driving limit for the probe part is 
achieved when it stimulates a two-photon 
transition while the coupler part is resonant 
with a single-photon one being strong enough to dress the 
system. To predict the frequencies of the sideband transitions that can appear for such A-T-like processes and to
model the experimentally observed splittings induced by the 
same field that probes them, we have built several self-consistent models in 
rotating frame and found a good 
agreement between the model, the experiment, and 
the numerical simulation. Interestingly, no new spectral lines appear when the system is tuned to the parameters where these A-T-like processes can occur. Instead, for example, one can see how the already present spectral 
lines $12/3$ (three-photon) and $10-02$ (single 
photon) smoothly morph to an additional avoided crossing of a 
non-standard size of $\frac{2\sqrt{3}}{3} 
\Omega_1$.

Another A-T-like effect may occur when both 
components of the driving operator are in the 
single-photon regime. Now, both transmons are 
dressed and probed simultaneously. We find that in this case 
the $11/2$ transition frequency is altered. When the SAAs are being tuned into resonance, this spectral line is being split into two hyperbolic curves. However, unlike the case of a standard avoided crossing, the doublet transition is never observed. The shape of this apparent splitting and its visibility 
depends qualitatively on the relation between the driving 
amplitudes; for instance, if they are equal, the splitting does not appear at all. For our sample, we have a fixed ratio 
of approximately two between $\Omega_1$ and 
$\Omega_2$ which still allows us to distinguish this effect experimentally. Notably, light dressing affects the frequency of the $11/2$ transition even far 
from the 01, 10 intersection. This means that a fast bSWAP 
gate should be performed at a different frequency 
than predicted by the unperturbed Hamiltonian if there is an asymmetry in the driving amplitudes $\Omega_{1,2}$. 

Interestingly, the self-consistent models for the 
observed effects imply that multi-photon 
processes may smoothly change their order. For 
example, for features II and III, we observe a 
continuous transformation of a three-photon and a 
single-photon processes to the second order in 
\autoref{eq:two-photon}, and of a two-photon 
transition to a zero-photon in 
\autoref{eq:zero-photon}. One could quantify this effect by evaluating the transition linewidths and measuring their power dependence for different coil currents.

Overall, irradiating an individual diatomic artificial molecule with intense light calls forth a plethora of effects that may be used to extend the validity of the well-known light-dressing models. These effects are much easier to find in human-made quantum devices than in natural coupled systems due to higher overall controllability and addressability of their parts. The relative ease in attaining the excitation powers that cause multiphoton transitions promises even more complex dynamics in multi-atom systems. We are looking forward to investigating strong interaction with light in larger artificial structures such as one- and two-dimensional arrays of SAA.

\section{Acknowledgements}

We gratefully acknowledge valuable discussions with I.S. Besedin, V.V. Ryazanov, and A.V. Ustinov. 
The experimental investigation was conducted with the support of Russian Science Foundation, Grant No. 16-12-00070. Numerical simulations were supported by the Russian Science Foundation (contract no. 16-12-00095). The theoretical work was supported by Ministry of Science and Higher Education of Russian Federation in the framework of Increase Competitiveness Program of the NUST MISIS (contract no. K2-2017-081). Devices/Samples were fabricated at the BMSTU Nanofabrication Facility (Functional Micro/Nanosystems, FMNS REC, ID 74300)

\appendix

\section{Full circuit model}\label{sec:full_model}

In this section, we substantiate our use of the simplified model in \autoref{Hsystem} and compare our experiment with previous works. From the scheme in \autoref{fig:experiment}~(b) and from \cite{yan2018tunable} we can conclude, that we have two types of interaction between SAAs and that our choice for $\hat H_{int}$ is valid. The only difference is in the representation of the charge operator $\hat n$ which we choose to be proportional to $\hat b^\dag + \hat b$ instead of $i(\hat b^\dag - \hat b)$ which is a question of the basis choice (the linear oscillator basis or the transmon eigenbasis). The flipped sign before the counter-RWA terms is important only for the coupling to the resonator. However, it can be corrected by an altered Schrieffer-Wolff transformation leading to identical final results. The symmetries of the drive and the coupling operator stay unchanged with this approach. Having established the correctness of our simplified model, we can proceed and calculate the coupling strengths.

The first one, $J_{12}$, is caused by the direct capacitive coupling, and following \cite{yan2018tunable} we see that:
\[
J_{12} = \frac{1}{2}(1 + \eta)\frac{C_{12}}{\sqrt{C_1 C_2}} \sqrt{\omega_1 \omega_2},
\] 
where $C_{12}$ is the mutual capacitance between the transmon islands, $C_{1,2}$ are their capacitances to the ground, and $\omega_{1,2}$ are defined as in the main text. The coefficient  $\eta$ is around 6$\cdot 10^{-2}$ for our parameters as will be shown later.

The second coupling mechanism is via the multimode quantum bus, similar to \cite{majer2007coupling, filipp2011multimode}. However, the models used in these papers can now be improved using \cite{yan2018tunable} (counter-RWA terms must be taken into account for higher modes that are in the strong dispersive regime) and \cite{malekakhlagh2017cutoff, gely2017convergence} that have solved the divergence problems and found rigorously the effective cut-off frequency $f_{max}$ that we will employ. For our case, $f_{max}$ can be calculated using the equivalent capacitance of the network at the open end of the $\lambda/4$ resonator:
\[
C_{eff} \approx C_{claw} + (C_{g1}^{-1} + C_{1}^{-1})^{-1} + (C_{g2}^{-1} + C_{2}^{-1})^{-1},
\] 
where $C_{g1, g2}$ are the capacitances between the transmons and the resonator, $C_{claw}$ is the direct capacitance of the claw coupler to the ground plane, and $C_{12}$ and capacitances to the drive antenna are neglected. Now, to find $f_{max}$ we need to solve the equation
\[
\left|\frac{1}{i 2\pi f_{max} C_{eff}}\right| = |Z_{res}| \approx 50\ \Omega
\]
which marks the frequency above which the capacitive impedance at the end of the resonator will become small enough (comparable to its wave impedance $Z_{res}$) to effectively change the boundary condition there to a short, turning off the electric field and thus the capacitive coupling of the higher-frequency modes to the transmons. 

Next, we use the expression for the coupling through the j\textsuperscript{th} mode with the frequency $\omega_j^r$ based on \cite{yan2018tunable} with the counter-RWA terms:

\begin{equation}
\label{eq:bus_coupling}
J^{(j)}_{bus} \approx \frac{g_1^{(j)}g_2^{(j)}}{2}\left(\frac{1}{\Delta_{1}^{(j)}}+\frac{1}{\Delta_{2}^{(j)}}-\frac{1}{\Sigma_{1}^{(j)}}-\frac{1}{\Sigma_2^{(j)}}\right),
\end{equation}
where $\Delta_{1, 2}^{(j)}=\omega_{1,2}-\omega_j^r < 0\ \forall j$, $\Sigma_{1,2}^{(j)}=\omega_{1,2}+\omega_j^r$, and
\begin{equation}
\label{eq:g}
	g_{1,2}^{(j)} =\frac{1}{2}\frac{C_{g1, g2}}{\sqrt{C_{1,2}C_r}}\sqrt{\omega_{1,2}\omega_j^r},
\end{equation}
where the effective mode capacitance $C_r = C_r' l_r/2 + C_{claw}$ does not depend on $j$ \cite{yan2018tunable, pozar2009microwave}. $C_r' = 160$ fF/mm being the per-unit-length capacitance may be calculated for a CPW (center width of 7 $\mu$m, gap of 4 $\mu$m) with standard means (here and below we take the substrate $\varepsilon$ to be 11.45 \cite{krupka2006measurements}); $l_r = 3.477$ mm is the resonator length without the claw. $C_{claw}$ may be found from EM simulations or from the equation determining the observed resonator frequency:
\begin{equation}
	\label{eq:resonator_freq}
	\omega_0 = \frac{1}{\sqrt{L_r^{(0)}(1+\alpha) (C_r' l_r/2 + C_{claw})}},
\end{equation}
where $\alpha = 0.14$ is the kinetic inductance contribution determined using two test resonators without claws and $L_r^{(0)} = 8 L_r' l_r/\pi^2$ is the geometric equivalent inductance of the fundamental mode, $L_r' = 0.4$ nH/mm.
From EM simulation, we find $C_{claw}$ = 64 fF; however, from \autoref{eq:resonator_freq} we obtain $C_{claw} = $ 53 fF. We are unable to explain why the difference is so significant, even though we could successfully describe the frequencies of all 4 different resonators on the chip (two with transmon pairs and claws, at 6.840 and 7.340 GHz, and two test ones, at 7.700 and 7.800 GHz) up to MHz accuracy with \autoref{eq:resonator_freq} while fitting only $\alpha$ and $C_{claw}$. Below we will use the smaller value for $C_{claw}$ determined from the experimental data.

\begin{figure}[t]
	\includegraphics[width=\linewidth]{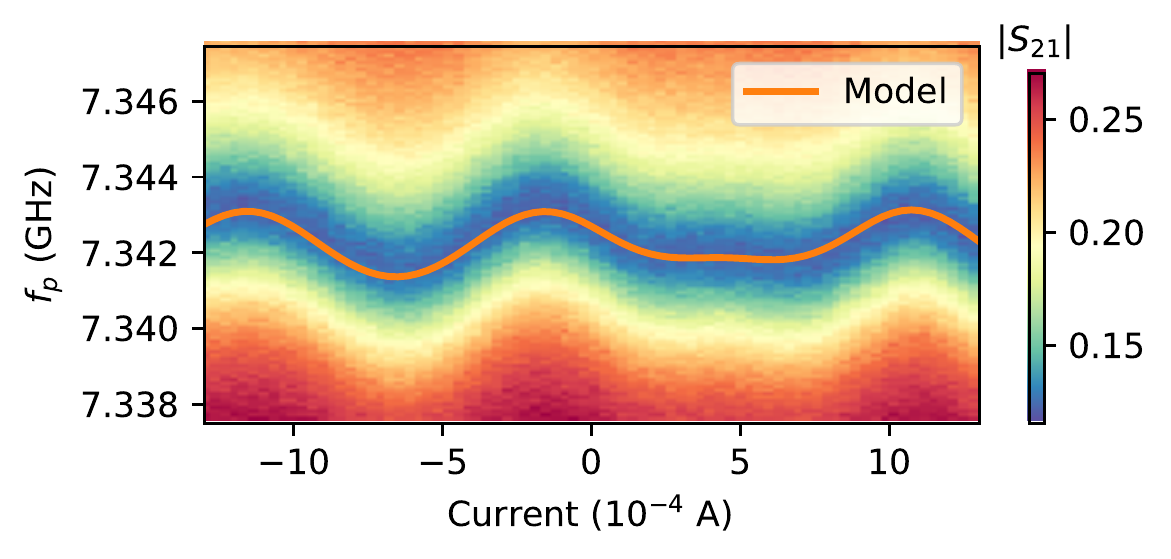}
	\caption{Determining the coupling strength via the resonator spectroscopy (external coil current is swept, flux bias lines turned off, transmission amplitude is shown). The transmon parameters $E_{J,\Sigma}^{(1,2)},\ E_C^{(1,2)},\ d^{(1,2)}$ are fixed to the values from Section \ref{sec:methods} A, and the bare cavity frequency, current sweet spots, current periods and coupling capacitance $C_g = C_{g1} = C_{g2}$ are the fitting parameters. The orange curve shows the theoretical prediction for the optimal values: $\omega_0/2\pi = 7.340$ GHz, sweet spots at -1.2 and 0.45 mA, periods of 1.15 and 0.74 mA, $C_g$ = 2.5 fF.}
	\label{fig:sts}
\end{figure}

Knowing	$C_r = 337$ fF, the capacitances $C_{g1, g2}$ may be determined with high accuracy by fitting the fundamental mode frequency dependence on the flux using \autoref{eq:g} when the transmon parameters are known (see also \cite{fedorov2019automated}). $C_{1,2}$ can be extracted from the experimental anharmonicity using the equation
\[
-\hbar \alpha = E_{C}^{1,2} = \frac{e^2}{2 C_{1,2}},
\]
which yields $C_{1,2}\approx$ 88 fF (we use here the renormalized $ E_{C}^{1,2} $ from \cite{gely2017convergence}, Eq. 6). The fit is shown in \autoref{fig:sts}, which yields the capacitances $C_{g1, g2}$ of 2.5 fF each, which agrees reasonably well with the value of 2.3 fF that we have obtained from EM simulation. We have checked that the sweet spots found this way agree with an independent direct SAM spectroscopy similar to \autoref{fig:two-tone}~(a). The discrepancy in $C_{g1,g2}$ probably arises from the real configuration of the ground electrode enclosing the chip in the sample holder which we do not include in the simulation. Similarly, we obtain simulated $C_{1,2}$ to be only 80 fF. This problem also hampers the accuracy of the calculated $C_{12}$ which unfortunately can not be found independently in this experiment. 

In overall, the total interaction 
\[
J = J_{12} + \sum_{j=0}^{j_{max}} J^{(j)}_{bus},
\]
where $j_{max}: \omega_j^r \approx 2\pi f_{max}$. For our configuration, the first term is positive, and the second is negative. Note that in contrast to \cite{filipp2011multimode}, the contribution from every mode is of the same negative sign since the transmons are located at the same resonator end and are below all modes in frequency. Below, we calculate the value of $J$ at $\omega_{1,2} \approx$ 5.3 GHz where we can determine its value experimentally from the size of the avoided crossing.

For the parameters $C_{1,2} = 88$ fF, $C_{g1, g2} = 2.5$ fF, $C_{claw} = 53$ fF we find that $C_{eff} \approx 59$ fF and $f_{max} \approx 55$ GHz. Since $\omega_j^r = (2j+1)\omega_0^r$, $\omega^r_0/2\pi \approx 7.3$ GHz, and $\omega^r_3/2\pi \approx 51 \text{ GHz} \approx f_{max}$, we only need to include 4 terms for the multimode virtual interaction between the transmon. We note that this very low cut-off is due to the large direct capacitance of the claw to the ground.

\begin{figure}[t]
	\includegraphics[width=\linewidth]{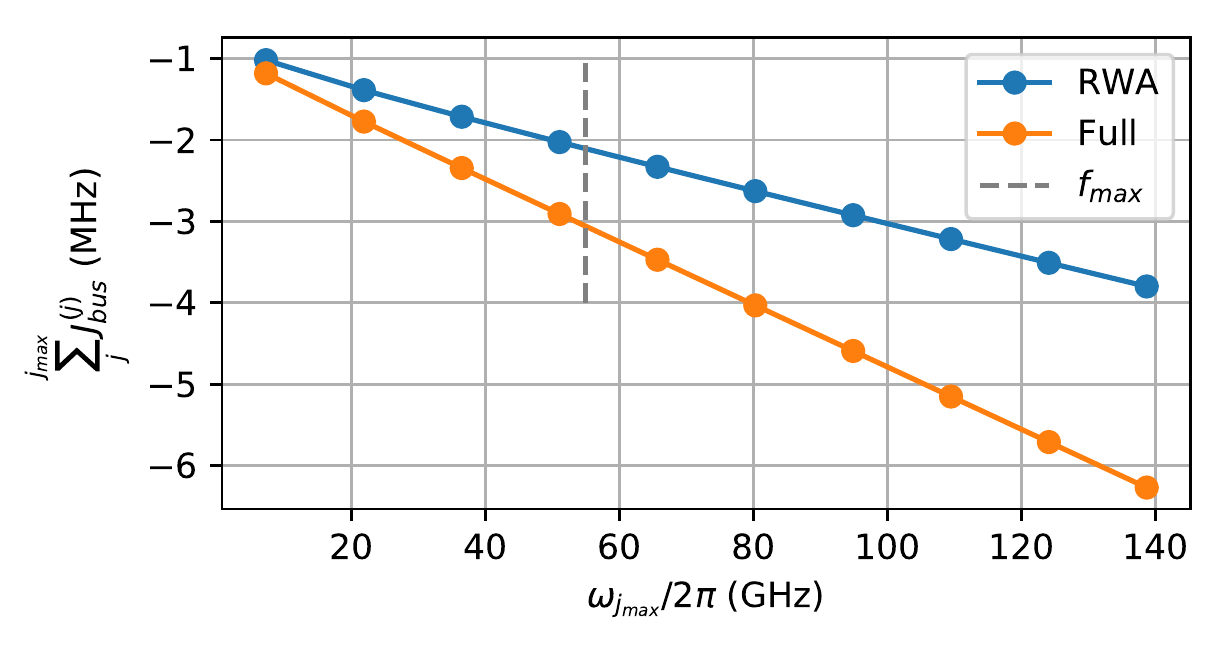}
	\caption{Multimode coupling vs $\omega_{j_{max}}$. Blue points show the RWA result (without the $\Sigma_{1,2}^{(j)}$ terms) and orange show the full calculation. Gray dashed line shows the calculated cut-off frequency of 55 GHz. The parameters for the calculation: $\omega_{1,2}/2\pi = 5.3$ GHz, $\omega_0/2\pi$ = 7.3 GHz, $C_{1,2} = $ 88 fF, $ C_r $ = 337 fF, $C_{g1,g2}$ = 2.5 fF giving $ g_{1,2}^{(0)}/2\pi = $ 45 MHz.}
	\label{fig:Jbus}
\end{figure}

Having established the required cut-off, we can evaluate the multimode contribution. In \autoref{fig:Jbus}, we show its value depending on the number of included modes with and without RWA. As one can see, the cumulative value at the cut-off is approximately $-3$ MHz for the non-RWA case and $-2$ MHz within RWA. For higher cut-off frequencies, the ratio between RWA and full solution tends to 2, since $\Sigma_{1,2}^{(j)} \approx -\Delta_{1,2}^{(j)}$ there.

Next, we find the direct coupling constant. As we have already mentioned, the capacitance $C_{12}$ can only be determined in a FEM simulation; however, it depends on the configuration of the ground plane stronger than $C_{g1, g2}$ and $C_{1,2}$ and thus is less accurately determined. The root of this is in the large size of the transmon island electrodes comparable to the substrate thickness of 0.4 mm. We use a sample holder with a machined cavity under the chip, and for a realistic FEM model, we obtain $C_{12} = 0.34$ fF. This yields $\eta = C_{g1}C_{g2}/C_{12}C_r = .059$, and the resulting direct coupling $J_{12} \approx 11$ MHz. 

Finally, adding both contributions, we find $J \approx 8$ MHz which agrees reasonably well with the experimental value of 8.69 MHz. The positive sign of the coupling is consistent with the location of the dark state in the lower branch of the avoided crossing between 01 and 10 in the experimental data which we have established in a separate measurement \cite{filipp2011multimode}. Despite the uncertainty in $J_{12}$, this result confirms that the cut-off frequency for cQED with a transmission line resonator should be calculated regarding the effective capacitance at its end and may be quite low. However, our accuracy is not enough to demonstrate confidently the necessity of the counter-RWA terms. It is notable that, in contrast to this work, \cite{filipp2011multimode} describes the experimental data within RWA, but does not adopt a possibly lower cut-off for their capacitances, so a separate study may be required to clarify the importance of the counter-rotating terms. Additionally, it would be interesting to compare the results with \cite{solgun2019simple}, presenting a classical approach to the problem, suitable for low-anharmonicity circuits.

To end this subsection, we would like to comment on the possible error sources in our reasoning. First, a more rigorous cut-off free analysis similar to \cite{gely2017convergence} should be conducted for our circuit to substantiate the use of the cut-off and find the renormalized Hamiltonian parameters. It would be possible if it is possible to find analytically the inverse capacitance matrix of the system, as was done in \cite{gely2017convergence}. Second, the use of the $(\hat b^\dag + \hat b) (\hat c^\dag + \hat c)$ form for the coupling via the bus is not absolutely rigorous, because for higher transmon levels \autoref{eq:bus_coupling} will include slightly different $\Delta_{1,2}^{(j)}$ and $\Sigma_{1,2}^{(j)}$ due to the non-zero $\alpha$ \cite{chow2010phd}.

\section{Standard Autler-Townes effect for a three-level $\Xi$ atom} \label{sec:3-level-at}

The underlying cause of the A-T effect is the dressing of the atomic levels by strong EM radiation. There are two equivalent mathematical models to describe it depending on whether the light is classical or quantized \cite{cohen1998atom}.

\begin{figure}[b]	
	\includegraphics[width=\linewidth]{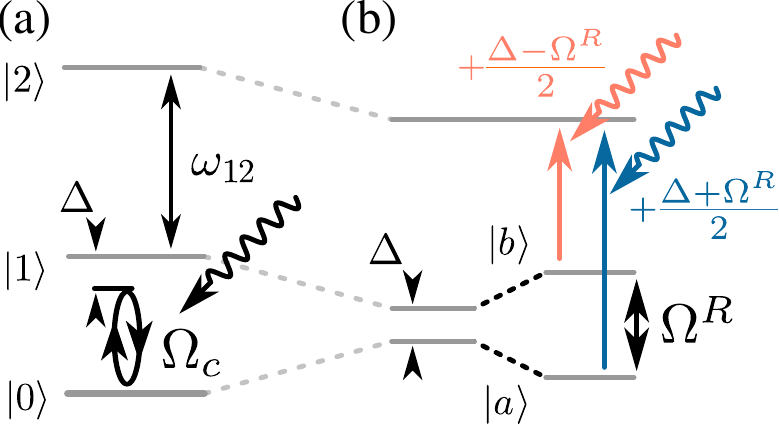}
	\caption{Illustrating the A-T splitting by  the classical driving in the rotating frame. \textbf{(a)} A three-level system is driven strongly by the coupler tone of amplitude $\Omega_c$ at frequency $\omega_c = \omega_{01}-\Delta$. \textbf{(b)} In the frame rotating with the drive, the $\ket{0}\rightarrow \ket{1}$ transition frequency changes to $\Delta$. However, when the RWA is applied and the Hamiltonian is re-diagonalized, the splitting between two lowest levels (dressed states $\ket{a}$ and $\ket{b}$) becomes $\hbar \Omega^R$. Now, a doublet transition from these levels to the state $\ket{2}$ at frequencies $\omega_{12}+(\Delta \pm \Omega^R)/2$ may be observed.} 
	\label{fig:at-standard}
\end{figure}

\paragraph{Classical derivation.} In the classical case, the mathematical description goes as follows. The $\ket{0}\rightarrow \ket{1}$ transition of frequency $\omega_{01}$ in a three-level system is driven strongly with an amplitude $\Omega_c$ at a detuning $\Delta$ (the \textit{coupler} tone). Additionally, weak radiation at $\omega_{p}$ and of amplitude $\Omega_p$ is sent at the $\ket{1} \rightarrow \ket{2}$, transition, of frequency $\omega_{12}$ (the \textit{probe} tone). This is illustrated in \autoref{fig:at-standard}~(a). The Hamiltonian for this driven system reads:

\begin{equation*}
\hat H_0/\hbar = \left[\begin{matrix}
0 & \Omega_{c} \cos{\left(\omega_{01}- \Delta \right)t} & 0\\
\Omega_{c} \cos{\left(\omega_{01} - \Delta \right)t}   & \omega_{01} &\Omega_{p} \cos\omega_p t  \\
0 & \Omega_{p} \cos{\omega_p t}  & \omega_{02}\end{matrix}\right],
\end{equation*}
where $\omega_{02} = \omega_{01} + \omega_{12}$.

Next, we move to the rotating frame by using an operator
\[
\hat R = \left[\begin{matrix} 
1 & 0 & 0\\0 & e^{- i t \left(\omega_{01}- \Delta\right)} & 0\\0 & 0 & e^{- i t \left(\omega_{01}- \Delta\right)}\end{matrix}\right].
\]
Note here that the state $\ket{2}$ is also rotated: this is convenient to preserve the frequency of the probe field. The new Hamiltonian is calculated as follows: $\hat H_1 = \hat R^\dag \hat H_0 \hat R - i \hat{R}^\dag \partial_t  \hat R$. The level structure without driving  is shown in the left part of \autoref{fig:at-standard}~(b). Applying the RWA, we obtain:
\begin{equation*}
\hat H_1/\hbar = \left[\begin{matrix}0 &\Omega_{c}/2 & 0\\\Omega_{c}/2 & \Delta & \Omega_{p} e^{i t \omega_p}/2\\0 & \Omega_{p} e^{-i t \omega_p}/2 & \omega_{02} - \omega_{01}+\Delta \end{matrix}\right].
\end{equation*}

Next, moving to the basis where the upper left 2x2 corner is diagonal (the right part of \autoref{fig:at-standard}~(b)), we obtain:

\begin{equation*}
\hat H_3/\hbar = \left[\begin{matrix} 
\frac{\Delta}{2} - \frac{\sqrt{\Delta^{2} + \Omega_{c}^{2}}}{2} & 
0 &
\frac{\Omega_{p} e^{i t \omega_{p}}}{2}\sin(\theta)
\\
0 & 
\frac{\Delta}{2} + \frac{\sqrt{\Delta^{2} + \Omega_{c}^{2}}}{2} & 
\frac{\Omega_{p} e^ {i t \omega_p}}{2}\cos(\theta)
\\
\frac{\Omega_{p} e^{-i t \omega_{p}} }{2}\sin(\theta) & 
\frac{\Omega_{p} e^{-i t \omega_{p}}}{2} \cos(\theta) & 
\omega_{12} +\Delta 
\end{matrix}\right].
\end{equation*}

One may see that the resonant conditions for the probe drive are now $\omega_p = \omega_{12} + {(\Delta \pm \Omega^R)}/{2}$, where $\Omega^R = \sqrt{\Omega_c^2 + \Delta^2}$, $\omega_{12} = \omega_{02}- \omega_{01}$, and its amplitude is renormalized by an angle $\theta,\ \tan 2\theta = -\Omega_c/\Delta$.

\paragraph{Quantum derivation.} For a fully quantum interpretation, the incident radiation at $\omega_{01}-\Delta$ is modelled as a single-mode quantum oscillator which is then coupled to the $\ket{0}\rightarrow \ket{1}$ transition. The Hamiltonian for the compound system reads:
\[
\begin{split}
\hat H_0/\hbar = \omega_{01} \ket{1}\bra{1} + \omega_{02} \ket{2}\bra{2} + (\omega_{01}-\Delta) \hat a^\dag \hat a + \\\ g (\hat a^\dag + \hat a)\otimes (\ket{1}\bra{0}+\ket{0}\bra{1}),
\end{split}
\]
where $g$ is the coupling strength and $a$ is the photon annihilation operator. After moving to the rotating frame with $\hat R = \exp[i t (\omega_{01}-\Delta)(\hat a^\dag \hat a + \ket{1}\bra{1}+\ket{2}\bra{2})]$ and applying the RWA, the Hamiltonian transforms into
\begin{equation}
\begin{split}
\hat H_1/\hbar = \Delta \ket{1}\bra{1} + (\omega_{12}+\Delta) \ket{2}\bra{2} + \\\ g \left[\hat a^\dag \otimes \ket{0}\bra{1} + \hat a \otimes \ket{1}\bra{0}\right].
\end{split}
\end{equation}
Now, presume that the resonator is in a coherent state $\ket{\alpha}$ with $\langle N\rangle = \alpha^2$ photons (it is time-independent in the rotating frame). Therefore, after tracing out the resonator subspace, from the interaction term we will obtain again the classical driving term $\Omega_c \left(\ket{0}\bra{1} + \ket{1}\bra{0}\right)$, where $\Omega_c = 2 g \sqrt{\langle N \rangle}$ in correspondence with the previous approach. The following steps completely reproduce the classical case if we add the probe tone $\Omega_p$ to the last equation.

\section{Degenerate perturbation theory}
\label{sec:dpt}

Since the Hamiltonian in the dressed basis is a 
useful illustration for 
\autoref{fig:main_scheme}, we provide it below in 
\autoref{eq:ham_matrix} for the case $\Delta_1 = 
0$, $\delta = \omega_d^{(2)} - 	\omega_d^{(1)} = 
\omega_d^{(2)} - \omega_1$. In the resonant case 
$2\Delta_2 + \alpha_2 = 0$, i.e. when the frame of the second transmon is rotated at its 
two-photon transition frequency (as in III), there is a degeneracy between 
states $\ket{a, 0}, \ket{a,2}$ and $\ket{b, 0}, 
\ket{b,2}$, while $\ket{a, 1}, \ket{b,1}$ are detuned from them by $|\alpha_2|/2$.

Note that in the dressed basis $\hat H_{int}$ that we treat as a perturbation has a sub- 
and super-diagonal block form and couples all the 
states $\ket{a, j}, \ket{a,j+1}$ and $\ket{b, j}, 
\ket{b,j+1}$.

The degenerate perturbation theory requires first of all to choose zero-order state vectors $\ket{N^0}$ from a degenerate subspace. The choice is arbitrary at the first glance because any linear combination of basis vectors $\ket{n}$ from this subspace will satisfy the unperturbed Schrödinger equation. However, if we demand the \textit{change} of $\ket{N^0}$ to be small under the perturbation $\hat V$, they become determined and are given by diagonalization of $\hat V$ in the degenerate subspace. Unfortunately, all matrix elements of $\hat V = \hat V_J$ are zero in both our degenerate subspaces, so technically any choice of zero-order states will diagonalize it. In other words, the degeneracy is not lifted in the first-order, and thus we have \cite{landau2013quantum} to diagonalize the matrix 
\begin{equation}
	M_{nn'} = \sum\limits_{m}\frac{V_{ nm}V_{mn^\prime}}{E_n-E_m}.
\end{equation}
Here, $\ket{n}$ and $\ket{n'}$ are the basis 
states from the degenerate subspace with energy 
$E_n$, and $V_{mn} = \bra{m} \hat  V \ket{n}$. 
The sum is over all other zero-order states 
$\ket{m}$ outside the degenerate subspace.

For example, in our case for one of the degenerate subspaces ($E_n^0 = -\Omega_1/2,\ \hat V = \hat V_J$) we obtain the following matrix:
\renewcommand{\kbldelim}{[}
\renewcommand{\kbrdelim}{]}
\begin{equation}
	M_{15} =\kbordermatrix{&\ket{a,0} & \ket{a, 2}\\
	\ket{a,0}&\frac{J^{2} \left(\Omega_{1} - \alpha_2\right)}{\alpha_2 \left(2 \Omega_{1} - \alpha_2\right)} &
	\frac{\sqrt{2} J^{2} \Omega_{1}}{\alpha_2 \left(2 \Omega_{1} - \alpha_2\right)}\\
	\ket{a, 2}&\frac{\sqrt{2} J^{2} \Omega_{1}}{\alpha_2 \left(2 \Omega_{1} - \alpha_2\right)} &
	\frac{2 J^{2} \left(\Omega_{1} - \alpha_2\right)}{\alpha_2 \left(2 \Omega_{1} - \alpha_2\right)}}.
\end{equation}
The normalized eigenvectors 
\begin{equation}
\ket{N} = C_n \ket{n} + C_{n'}\ket{n'}, 
\ket{N'} = C'_{n}\ket{n} + C'_{n'}\ket{n'}
\end{equation}
of the matrix are the desired zero-order superpositions. Next, we first-order correct them as usual:
\begin{equation}
	\ket{N}+\ket{N^{1}} = \sum_{j=n,n'} C_{j}\ket{j} + \sum_{j\ne n, n'}\ket{j}\frac{\bra{j}\hat V\ket{N}}{E_n-E_j}.
\end{equation}

Similarly, the first-order correction for a non-degenerate state is
\begin{equation}
	\ket{m^1} = \sum_{j\ne m}\ket{m}\frac{\bra{j}\hat V\ket{m}}{E_m-E_j}.
\end{equation}

\renewcommand{\kbldelim}{[}
\renewcommand{\kbrdelim}{]}
\begin{widetext}
	\begin{equation}
	\hat H^D + \hat V_J + \hat V_t(t)  = 
	\kbordermatrix{
		&\ket{a,0} & \ket{b,0} & \ket{a,1} & 
		\ket{b,1} & \ket{a,2} & \ket{b,2} \\
		\ket{a,0}& -\frac{\Omega_1}{2} & 0 & 
		\frac{\Omega_{2}e^{i \text{$\delta$} 
				t}}{2} -\frac{J}{2} & \frac{J}{2} & 0 & 0 
		\\
		\ket{b,0} & 0 & \frac{\Omega_1}{2} & 
		-\frac{J}{2} & \frac{\Omega_2 e^{i \delta 
				t}}{2} + \frac{J}{2} & 0 & 0 \\
		\ket{a,1} &\frac{\Omega_2 e^{-i 
				\text{$\delta$} t} }{2}  - \frac{J}{2} & 
		-\frac{J}{2} & - \frac{\alpha_2}{2} - 
		\frac{\Omega_1}{2}&
		0 & \frac{\text{$\Omega_2$} e^{i 
				\text{$\delta$} t}}{\sqrt{2}} 
		-\frac{J}{\sqrt2} & \frac{J}{\sqrt2} \\
		\ket{b,1} & \frac{J}{2} & 
		\frac{\text{$\Omega_2$} e^{-i 
				\text{$\delta $} t}}{2} +\frac{J}{2} & 0 
		& - \frac{\alpha_2}{2} + \frac{\Omega_1}{2} 
		& -\frac{J}{\sqrt 2} & 
		\frac{\text{$\Omega_2$} e^{i \text{$\delta
					$} t}}{\sqrt 2} + 
		\frac{J}{\sqrt 2} \\
		\ket{a,2} & 0 & 0 & \frac{\Omega_2 e^{-i 
				\delta t}}{\sqrt 2} - \frac{J}{\sqrt 2} & 
		-\frac{J}{\sqrt 2} &
		-\frac{\Omega_{1}}{2} & 0 \\
		\ket{b,2} & 0 & 0 & \frac{J}{\sqrt 2} & 
		\frac{\text{$\Omega_2$} e^{-i 
				\text{$\delta $} t}}{\sqrt 2} + 
		\frac{J}{\sqrt 2} & 0 & 
		\frac{\Omega_{1}}{2}\\
	}
	\label{eq:ham_matrix}
	\end{equation}
\end{widetext}

\section{Transition rates of the two-photon process}\label{sec:2pp}
To quantify the visibility of the sideband transitions depending on the coupling strengths, we will employ the time-dependent perturbation theory that gives analytical expressions for the transition rates for single and multi-photon processes following \cite{faisal2013theory}.

Let us consider a time-dependent perturbation $\hat V(t) =\frac{\hbar\Omega_{2}}{2}(\hat{c}e^{i\omega_d t}+\hat{c}^{\dagger}e^{-i\omega_d t}) $ to an unperturbed Hamiltonian $\hat H$. In the interaction picture, the Schrödinger equation reads
$$
i\hbar \partial_t{\psi(t)} = \hat {V_I}(t)\psi(t),
$$ 
where 
$$
\hat {V}_I(t) = e^{\frac{it}{\hbar}\hat H}\hat V(t)e^{-\frac{it}{\hbar}\hat H}.
$$
The eigenstate of $\hat H$ at energy $E_j$ will be denoted as $\ket{j}$ whence it follows that
$$
\bra{j}\hat{V}_I(t)\ket{i} = e^{i\omega_{ij} t}\bra{j}\hat V(t)\ket{i},
$$
where $\hbar \omega_{ij} = E_j - E_i$. Formally, we can write down the solution of the Schrödinger equation in the interaction picture corresponding to the initial state $\ket{i}$ as
\begin{equation}\label{forpsi}
	\ket{\psi_i(t)}=\ket{i} - \frac{i}{\hbar} \int_{-\infty}^{t} d\tau_1 \hat{V}_I(\tau_1)\ket{{\psi_i}(\tau_1)}.
\end{equation} 
Solving \autoref{forpsi} by simple iterations 
gives the series solution for it:
\begin{align}
	U(t,-\infty) &= 1 + \sum_{n} U^{(n)}(t,-\infty),\\
	U^{(n)}(t,-\infty) &= \left(-\frac{i}{\hbar}\right)^n \int\limits_{-\infty}^{t}d \tau_1\hat V_I(\tau_1)...\int\limits_{-\infty}^{\tau_{n-1}}d \tau_n\hat{V}_I(\tau_n),
\end{align}
where $U(t,-\infty)$ is the evolution operator. For a weak perturbation, this series may be truncated at a finite term $n$, and the $n^{\text{th}}$ order transition amplitude $\bra{f}U^{(n)}(t,-\infty)\ket{i}$ can be evaluated:
\begin{equation}
	\begin{aligned}
	\bra{f}U^{(1)}(t,-\infty)\ket{i}&=-\frac{i}{\hbar}\int\limits_{-\infty}^{t}d\tau_1\hat{V}_I(\tau_1)\\ &=-\frac{i\Omega_2}{2}\big[\int\limits_{-\infty}^{t}d\tau_1 e^{i(\omega_{if}-\omega_d)\tau_1}\bra{f}\hat{c}^{\dagger}\ket{i}+\\
	&\quad\quad\quad\quad\int\limits_{-\infty}^{t}d\tau_1 e^{i(\omega_{if}+\omega_d)\tau_1}\bra{f}\hat{c}\ket{i}\big].
	\end{aligned}
\end{equation}
\begin{equation}
	\begin{aligned}
	\bra{f}U^{(2)}(t,-\infty)\ket{i}=-\frac{1}{\hbar^2}\bra{f}\int\limits_{-\infty}^{t}d\tau_1\hat{V}_I(\tau_1)\int\limits_{-\infty}^{\tau_1}d\tau_2\hat{V}_I(\tau_2)\ket{i}=\\
	-\frac{1}{\hbar^2}\sum_j\int\limits_{-\infty}^{t}d\tau_1 e^{i\omega_{jf}\tau_1}\bra{f}\hat V(\tau_1)\ket{j}\int\limits_{-\infty}^{\tau_1}d\tau_2 e^{i\omega_{ij}\tau_2}\bra{j}\hat V(\tau_2)\ket{i}.
	\end{aligned}
\end{equation}
Going over the long-time limit $t\rightarrow\infty$ in the final integral and collecting the terms belonging to the same delta functions gives \cite{faisal2013theory}
\begin{equation}
	\begin{split}
	\bra{f}U^{(2)}(t, -\infty)\ket{i}=\\\frac{2\pi i\Omega_2^2}{4}\bigg[\delta(\omega_{if}+2\omega_d)\sum_j\frac{\bra{f}\hat{c}\ket{j}\bra{j}\hat{c}\ket{i}}{\omega_{ij}+\omega_d}+\\
	\delta(\omega_{if})\bigg(\sum_j\frac{\bra{f}\hat{c}^{\dagger}\ket{j}\bra{j}\hat{c}\ket{i}}{\omega_{ij}+\omega_d}+\\
	\sum_j\frac{\bra{f}\hat{c}\ket{j}\bra{j}\hat{c}^{\dagger}\ket{i}}{\omega_{ij}-\omega_d}\bigg)+\\
	\delta(\omega_{if}-2\omega_d)\sum_j\frac{\bra{f}\hat{c}^{\dagger}\ket{j}\bra{j}\hat{c}^{\dagger}\ket{i}}{\omega_{ij}-\omega_d}\bigg].
	\end{split}
\end{equation}
The probability of the two-photon transition from $\ket{i}$ to $\ket{f}$ is the square modulus of the corresponding amplitude, $P^{(2)}_{i\rightarrow f}= |\bra{f}\hat U^{(2)}(\infty, -\infty)\ket{i}|^2$.
The rate (probability per unit interaction time $T$) of absorption of $2$ photons is defined as
\begin{equation}
	W^{(2)}_{i\rightarrow f}=\lim\limits_{T\rightarrow\infty}\frac{P^{(2)}_{i\rightarrow f}}{T}.
\end{equation}
To cancel out one of the delta functions in the resulting expression, we can use the identity 
\begin{equation}\nonumber
	\delta(\omega_f-\omega_i-2\omega_d) =\frac{1}{2\pi} \lim\limits_{T\rightarrow\infty}\int_{-T/2}^{T/2}e^{i(\omega_{if}-2\omega_d)t}dt = \lim\limits_{T\rightarrow\infty}\frac{T}{2\pi}.
\end{equation} 

For a two-photon emission:
\begin{equation}\label{prob}
\begin{aligned}
	W^{(2)}_{i\rightarrow f}&=\frac{\pi\Omega_2^4}{8}\bigg|\sum_j\frac{\bra{f}\hat{c}\ket{j}\bra{j}\hat{c}\ket{i}}{\omega_{ij}+\omega_d}\bigg|^2\delta(\omega_{if}+2\omega_d)\\ 
	&=R_{i\rightarrow f}^{(2)}\delta(\omega_{if}+2\omega_d).
\end{aligned}
\end{equation}

For a two-photon absorption:
\begin{equation}\label{prob1}
\begin{aligned}
W^{(2)}_{f\rightarrow i}&=\frac{\pi\Omega_2^4}{8}\bigg|\sum_j\frac{\bra{f}\hat{c}^{\dagger}\ket{j}\bra{j}\hat{c}^{\dagger}\ket{i}}{\omega_{ij}-\omega_d}\bigg|^2\delta(\omega_{if}-2\omega_d)\\ 
&=R_{f\rightarrow i}^{(2)}\delta(\omega_{if}-2\omega_d).
\end{aligned}
\end{equation}

For our case, $ \omega_d = \delta$ which leads to \eqref{eq:rates}.

\section{Measurement setup and 
methods}\label{sec:meas_setup}
The sample was measured in a BlueFors LD250 dilution refrigerator at 16 mK. For the readout a Keysight PNA-L N5232A VNA was used. For the coherent excitation of the SAM, we used an Agilent MXG N5183B analog signal generator. The sample was 
flux biased using Yokogawa GS200 current sources (two for the flux lines and one for the external coil wrapped around the sample holder).

Input microwave lines were isolated from 
the high-temperature noise with 60 dB of attenuation 
(10 @ 4K, 10 @ 1K, 20 @ 100 and 16 mK) and 
custom IR filters. The effective on-chip 
attenuation between the drive line and the 
transmons was calculated in Sonnet to be around 
70 dB @ 6 GHz. Coaxial flux-bias lines were 
attenuated by 20 dB @ 4K and IR filtered as well. 
Output path contained two 20 dB isolators and two 
amplifiers: a 4-14 GHz LNF amplifier at the 4 K 
stage and a room-temperature LNF amplifier.

As the main experimental method, we have employed the so-called two-tone spectroscopy which consists of exciting the SAM with monochromatic light at a certain frequency (first tone) until the steady state is reached while simultaneously measuring the signal transmission at the readout resonator frequency (second tone). This technique yields the average value of the joint measurement operator $\hat M$ in the steady-state. To find the resonator frequency, we have been fitting its complex $S_{21}$ response for each current with the second tone turned off with a method similar to the one described in our prior work \cite{fedorov2019automated} which employs the \textit{circlefit} library \cite{probst2015efficient}; the source code of our measurement script may be found at GitHub \footnote{\url{https://github.com/vdrhtc/Measurement-automation/blob/master/lib2/FastTwoToneSpectroscopy.py}}.

\section{Sample fabrication}\label{sec:fab}
The device was fabricated on a high-resistivity
Si wafer (10 kOhm$\cdot$cm). First, the wafer 
was cleaned with Pirahna, HF, and then coated 
with bilayer MMA/PMMA resist stack. The nominal 
after-bake thickness of the MAA and PMMA are 800 
and 100 nm, respectively. The
bilayer resist stack was exposed using a 50kV 
Raith Voyager EBL system and then developed. 
Next, the sample was placed into a high-vacuum 
electron-beam evaporation
chamber (Plassys) and after a gentle ion-milling 
step, a double-angle evaporation technique at 10 degrees 
was used to deposit Al/AlOx/Al layer (25/45 nm). Finally, hot 
NMP followed by IPA was used to lift off the 
resist mask stack.

\bibliography{papers_bibliography}
\end{document}